\documentclass[iop,appendixfloats]{emulateapj}

\usepackage[usenames,dvipsnames]{color}
\usepackage{color}
\usepackage{amsmath}
\usepackage{float}
\usepackage{ctable}
\usepackage{refcount}
\usepackage{lineno}
\newcommand{\figurepath}{.}

\newbox\grsign \setbox\grsign=\hbox{$>$}
\newdimen\grdimen \grdimen=\ht\grsign
\newbox\laxbox \newbox\gaxbox
\setbox\gaxbox=\hbox{\raise.5ex\hbox{$>$}\llap
     {\lower.5ex\hbox{$\sim$}}}\ht1=\grdimen\dp1=0pt
\setbox\laxbox=\hbox{\raise.5ex\hbox{$<$}\llap
     {\lower.5ex\hbox{$\sim$}}}\ht2=\grdimen\dp2=0pt

\shorttitle{Magnetized jet formation in a binary star}
\shortauthors{Sheikhnezami \& Sepahvand}
\begin{document}
\title{Exploring the Jet Formation in binary systems applying 3D MHD simulations}

\author{Somayeh Sheikhnezami
\altaffilmark{1,2}
\&
        Mostafa Sepahvand
\altaffilmark{1}
}
\altaffiltext{1}{Department of Physics, Institute for Advanced Studies in Basic Sciences (IASBS), P.O. Box 11365-9161, Zanjan, Iran}
\altaffiltext{2}{School of Astronomy, Institute for Research in Fundamental Sciences (IPM), Tehran, 19395-5746, Iran}
\email{snezami@iasbs.ac.ir}

     \date{\today}
\begin{abstract}

  We investigate the formation of an ideal magnetized jet that originates from a disk acting as a boundary by conducting axisymmetric MHD simulations. Our simulations demonstrate that the magnetized jet is consistently launched and reaches a stable state. We extended the model setup to three dimensions to further advance our study. We performed 3D MHD simulations of the jet launched from a disk surface, achieving a stable and appropriate model setup. Additionally, we expanded our study by incorporating the companion star and examining the influence of the Roche potential on the jet material. Specifically, we investigate whether including the companion star in the model significantly affects the dynamical evolution of the jet. Our findings reveal the formation of an ``arc-like'' structure in the density map of the jet cross-section, which is attributed to the direct tidal effects. This implies that while the primary physical effects and characteristics of the outflow on a larger scale are attributed to the host accretion disk, the direct tidal effects on the jet dynamics have a substantial impact, particularly in the vicinity of the Roche lobe and towards the secondary star.

\end{abstract}
\section{Introduction}

Astrophysical jets are fast-moving streams of material that have been observed in various astrophysical systems, including young stellar objects (YSOs), microquasars, and active galactic nuclei (AGN). They transfer energy, angular momentum, and mass to larger scales, such as molecular clouds, and significantly impact environmental evolution and processes like star formation. 
Jets are a well-known phenomenon, and their propagation and interaction with the interstellar medium have been the subject of numerous studies for many years\citep{2007ApJ...668.1028B,2010MNRAS.402....7M,2012A&A...539A..57P,2014MNRAS.439.2903C,2019ApJ...883..160S}. Furthermore, jets and outflows are observable phenomena that can be used to trace the physical properties of their host sources, such as the central object and underlying accretion disk.
 It is now well known that the large-scale magnetic field and its interaction with the host accretion disk play a major role in jet formation and acceleration \citep{1982MNRAS.199..883B,1983ApJ...274..677P,1985PASJ...37...31S,1997A&A...319..340F}. As a result, many studies have applied numerical simulations to investigate jet launching and acceleration while considering the disk as a boundary \citep{1997ApJ...482..712O,2006ApJ...653L..33A,2011ApJ...742...56V, 2011ApJ...737...42P,2013MNRAS.429.2482P,2022ApJ...933...71F} or including the disk evolution \citep{2002ApJ...581..988C,2007A&A...469..811Z,2010A&A...512A..82M,2012ApJ...757...65S,2013ApJ...774...12F,Stepanovs2014,2020ApJ...900...60M}.
There are plenty of studies of astrophysical jets that have focused on jet formation in single stellar systems. However, there are significant observations of jet precession or curved ballistic motion that strongly suggest that the jet source is part of a binary or even a multiple system. 
Numerous studies investigating binary systems including jets have been performed utilizing observational data or simulation techniques \citep{Fendt1998, 2000ApJ...535..833S, 2002MNRAS.335.1100C, 2004HEAD....8.2903M, 2007A&A...476L..17A, 2009ApJ...692..943C,2014xru..confE.147M, 2014cosp...40E1454K, 2016A&A...593A.132P, Beltran2016, 2019ASSP...55...71M, 2019A&A...622L...3E, 2019IAUS..346...34M, 2021MNRAS.503..704M, 2021MNRAS.503.3145B,2021MNRAS.503.5448D,2022A&A...666A.105P,2022A&A...661A.117L}.

Recently, numerical simulations of the disk forming the jet have been developed in three dimensions, making it feasible to study the non-axisymmetric properties of the jet as part of a binary system \citep{2015ApJ...814..113S}. Moreover, in our recent papers, we have studied the effects of including the companion star in the model setup \citep{2018ApJ...861...11S,2022ApJ...925..161S}. One important result from our 3D numerical MHD simulations is the observation of spiral structures in both the disk and the jet. \cite{2022ApJ...925..161S} have discussed and demonstrated that the spiral structure observed in the jet is injected from the disk into the outflow and then propagates along the jet. This suggests that the main structures seen in the jet are delivered from the underlying disk.
However, at low altitudes where the disk wind is close to the disk, the tidal forces on the jet materials are expected to be significant. As a result, the flow of the jet across the Roche lobe can be directly influenced by these tidal forces, leaving an imprint on the jet dynamics. Consequently, the resulting structure inside the jet is not solely the outcome of the injection process.

To investigate the direct impacts of tidal forces on the evolution of the jet within a binary star system, our specific focus is on the dynamical evolution of the jet. In this context, we treat the disk as a boundary. Initially, we utilize an axisymmetric simulation to explore and determine an appropriate model setup that enables the effective launching of a fast and stable jet from the disk boundary.
Subsequently, we advance our setup to ``three dimensions``, enabling us to conduct a comprehensive study of jet formation following the full 3D evolution of a jet launched in single star and binary star systems. This allows us to examine the effects of the secondary star on the dynamics of the jet within binary systems. In our 3D simulations, the disk acts as a surface, and we do not consider its evolution.

{ Our paper is organized as follows: In Section 2, we discuss our model setup. In section 3 , we present and discuss our 3D simulation of a jet launched in a single star system. In section 4, we present the 3D simulations of the jet formation in a binary system, and we compare the obtained results. In Section 5, we summarize the results.}

\section{Model setup}

We investigate the launching of an MHD outflow from a sub-Keplerian disk, which serves as a boundary. Our primary focus is to examine the direct influence of a companion star on the structure and evolution of the outflow. This is achieved by incorporating the Roche potential and conducting 3D MHD simulation runs.

For all simulations presented in this study, we utilize version 4.4 of the MHD PLUTO code \citep{2007ApJS..170..228M, 2012ApJS..198....7M}. This code allows us to solve the time-dependent Ideal MHD equations, which include the conservation of mass, momentum, and energy,

\begin{equation}
	\frac{\partial \rho}{\partial t}+(\vec{v} \cdot \vec{\nabla}) \rho+\rho \vec{\nabla} \cdot \vec{v}=0.
\end{equation}

\begin{equation}
\rho\left(\frac{\partial {\vec{v}}}{\partial \mathrm{t}}+({\vec{v}} \cdot {\vec \nabla}) \vec{v}\right)+{\vec \nabla} \mathrm{P}+\frac{1}{4 \pi}({\vec \nabla} \times \vec{B}) \times \vec{B}+\rho \vec{\nabla}  \Phi=0.
\end{equation}

\begin{equation}
\frac{\partial(\rho \mathrm{E})}{\partial \mathrm{t}}+{\vec \nabla} \cdot\left[\rho E \vec{v}+\left(\mathrm{P}+\frac{\mathrm{B}^2}{8 \pi}\right) \vec{v}\right]-\vec{B}(\vec{v} \cdot \vec{B})+\rho({\vec \nabla} \Phi) \cdot \vec{v}=0.
\end{equation}

Here, $\rho$ is the mass density, $\vec v$ is the velocity, $P$ is the thermal gas pressure,
$\vec B$ represents the magnetic field and $\Phi$ indicates the gravitational potential. To describe the evolution of the magnetic field the induction equation is applied, i.e.,
\begin{equation}
	\frac{\partial \vec{B}}{\partial \mathrm{t}}-{\vec \nabla} \times(\vec{v} \times \vec{B})=0.
\end{equation}

The gas pressure follows an equation of state $ \mathrm{P}=(\gamma-1) \mathrm{u} $
with the polytropic index $\gamma$ and the internal energy density $u$.
The total energy density is,

\begin{equation}
	e=\frac{\mathrm{p}}{\gamma-1}+\frac{\mathrm{\rho v}^2}{2}+\frac{\mathrm{B}^2}{8 \pi}+\rho\Phi.
\end{equation}

In this study, the heating and cooling processes are not considered to avoid the complexity. 
\begin{table}
\caption{ Characteristic simulation parameters:
The input parameters for our simulation runs are defined in this table. The parameters shown in the table are plasma beta 
$\beta_p=\frac{8 \pi P}{B^2}$, the mass load coefficient
$\eta=\frac{\rho V_P}{B_P}$, the binary separation $D$, the orbital period of the binary $T$ and the mass ratio of the secondary to the primary star is $ q= \frac{M_s}{M_p} $.}
\begin{center}
\begin{tabular}{|c|c|c|c|c|c|}
\hline
\hline
\noalign{\smallskip}

Run  & ${\beta_{p}}$ & ${\eta}$ & D  & $q$ & T \\
\noalign{\smallskip}
\hline
\noalign{\smallskip}
\noalign{\smallskip}
{\rm 2Dc0} \footnote{The first row is the axisymmetric run and the rests are 3D runs.}       & 20  & 0.001  & 0    & 0  & -    \\
\noalign{\smallskip}
\hline
\noalign{\smallskip}
 {\rm 3Dc0}      &  20  &  0.001  &  0    &   0 &-     \\
 {\rm 3Dc1}       &  20  &  0.001  &  100  &   0.1& 5987.74    \\
 {\rm 3Dc2}       &  20  &  0.001  &  120  &   1.0 & 5837.36   \\
 {\rm 3Dc3}       &  20  &  0.001  &  100  &   2.0 & 3625.75   \\
\noalign{\smallskip}
 \hline
 \noalign{\smallskip}
 \end{tabular}
 \end{center}
\label{tbl:0}
\end{table}

\begin{figure*}
\centering
\includegraphics[width=14cm]{\figurepath/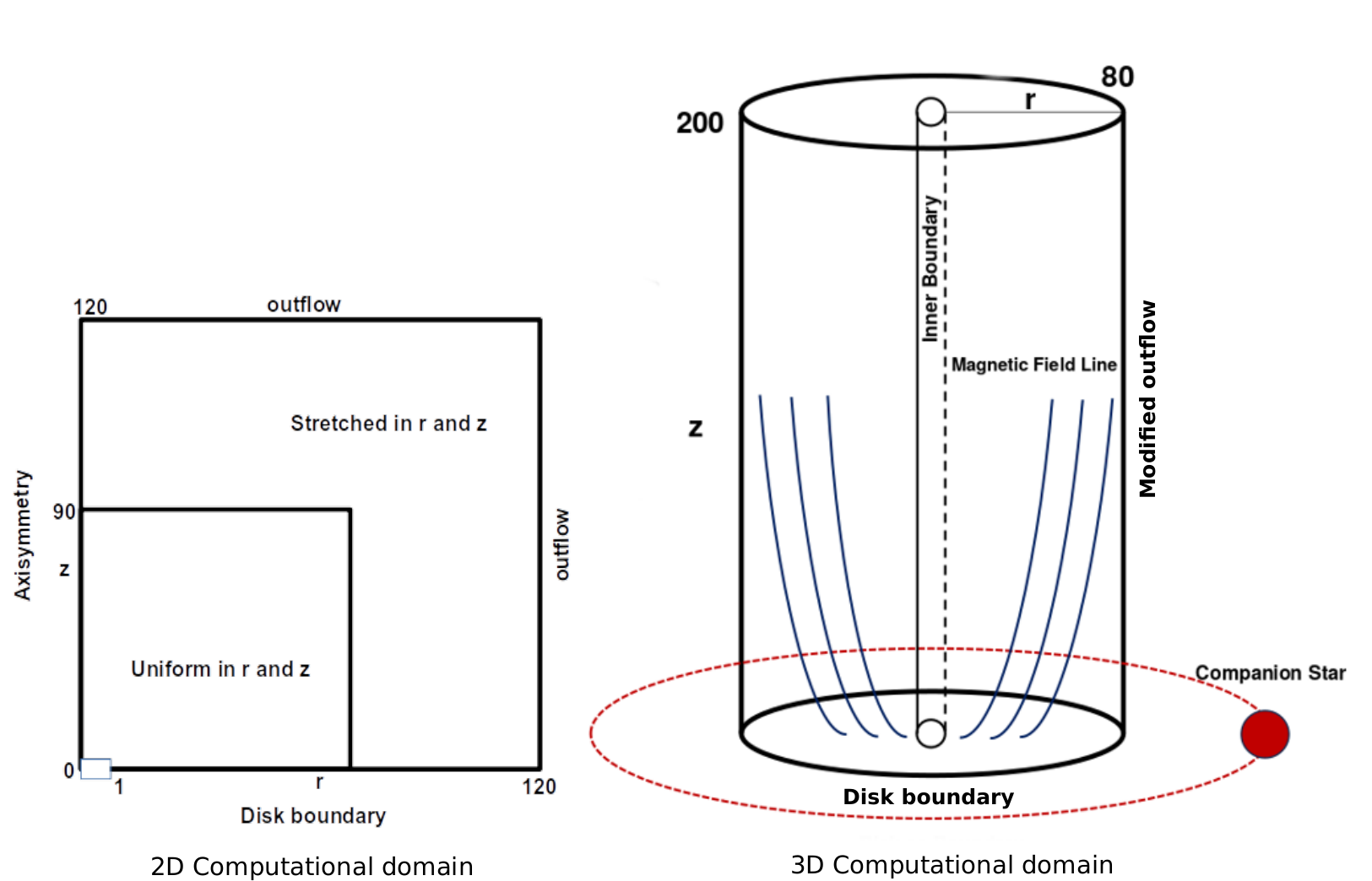}
\caption{ Computational domain. The left panel shows our axisymmetric computational domain and the corresponding boundary conditions. The right panel, on the other hand, provides a schematic representation of the 3D setup, which includes the boundary conditions and the orbital plane of the binary star.}
\label{fig:boundary-condition}
\end{figure*}

\subsection{Units and normalization}

In this section, we explain the normalization we use throughout the paper. The inner radius of the disk boundary is shown by $r_{\rm i}$, and we adopt $r_{\rm i}=1$.
All distances are expressed in units of inner disk radius $r_{\rm i}$.
The gas pressure and density at this radius are $P_{\rm i}$ and  $\rho_{\rm i}$.
Velocities are expressed in units of the Keplerian velocity at the inner radius of the disk
$v_{\rm k, i}$ which we adopt $v_{\rm k,i = 1}$. 
Time is estimated in units of 
$t_{\rm i}=\frac{r_{\rm i}}{v_{\rm k,i}}$.
The magnetic field is measured in units of the magnetic field at the inner disk radius $B_{\rm i}$
which can be defined as,
$B_{\rm i}=\sqrt{\frac{2 P_{\rm i} }{\beta_{\rm p} } }$
where ${\beta_{\rm p} }$ denotes the plasma beta parameter of the materials\footnote{In PLUTO code the magnetic field is normalized considering $4\pi =1$ footnote\label{MyFootNoteLabel}}.

\subsection{Initial Conditions}

We define the initial conditions following the setup applied by  \citet{2011ApJ...742...56V} to explore a magnetized jet emanating from a disk boundary prescribed in sub-Keplerian rotation.
For our axisymmetric reference run, we define the potential due to the gravity of the central star, denoted by $\Phi$, using the following equation:

\begin{equation}
    \Phi_{(\mathrm{r}, \mathrm{z})} = GM_{\odot}\left(\left(\mathrm{r}+\mathrm{r}_{\mathrm{g}}\right)^2+\left(\mathrm{z}+\mathrm{z}_{\mathrm{g}}\right)^2\right) .
\end{equation}

We implement the companion star by applying the Roche potential (see section \ref{3D binary}).
Here, $ (\mathrm{r}_{\mathrm{g}},\mathrm{z}_{\mathrm{g}}) $
are softening parameters included to avoid singularity at the origin and are taken as
 $ \mathrm{r}_{\mathrm{g}}=\mathrm{z}_{\mathrm{g}}=0.21 $
in our setup.
Initially, we prescribe a hydrostatic equilibrium with a density distribution,

 \begin{equation}
\rho_{(r,z)}=\rho_{\rm i}\left(\left(r+r_g\right)^2+\left(z+z_g\right)^2\right)^{-\frac{3}{4}}
 \end{equation}

The thermal pressure follows the polytropic equation of states
$ p =\mathrm{K} \mathrm{p}^\gamma$ where $\gamma=5/3$ is the
polytropic index.
We set the initial radial and vertical velocities to be proportional to the sound speed at the radius of the inner disk boundary, denoted by $C_{\rm s}= \sqrt{\gamma p_{\rm i} / \rho_{\rm i} }$. These values are arbitrary.

The initial magnetic field is initially set to be purely poloidal and follows the prescription used in \citet{2011ApJ...742...56V}. The vector potential, denoted as $\mathrm{A}_{\Phi}$, is defined as follows:

\begin{equation}
\mathrm{A}{\phi}=\frac{\left(\sqrt{\mathrm{r}^2+\mathrm{(z+z{\rm d}})^2}-\mathrm{(z+z_{\rm d})}\right)}{\mathrm{r}},
\end{equation}

where the parameter $z_{\rm d}$ represents the thickness of the disk.

Additionally, we have also investigated an alternative initial magnetic field distribution defined by \cite{2007A&A...469..811Z}. In this case, the magnetic flux function $\psi$ is used and given by:

\begin{equation}
\displaystyle
\psi(r,z) = \frac{3}{4} B_{z,i} r_{\rm i}^2
\left(\frac{r}{r_{\rm i}}\right)^{3/4}
\frac{m^{5/4}}{\left( m^2 + \left(z/r\right)^2\right)^{5/8} },
\label{eq;magpsi}
\end{equation}

where $B_{z,i}$ represents the vertical field strength at $(r=r_{\rm i},z=0)$. The vector potential, denoted as $A_{\rm \phi}$, is then given by:
\begin{equation}
A_{\rm \phi}= \frac{\psi}{r}.
\end{equation}

The latter definition is more consistent with our 3D computational grids which excludes the innermost part of the radial axis $(r< 0.5)$. As a result, we do not encounter any cuts in our magnetic field distribution.

\subsection{Numerical setup}

In our axisymmetric model setup, we use a numerical grid of $ (600\times600) $  cells applying cylindrical coordinates with domain size of $(\mathrm{r} \times \mathrm{z})=(120 \times 120) \mathrm{r}_{\rm {i}} $.
The computational domain covers the rectangular grids $(400\times400)$ cells with uniform spacing in the radial and vertical direction for the domain size of $ (\mathrm{r} \times \mathrm{z})=(20 \times 20) \mathrm{r}_{\rm{i}} $ and applying extra stretched grids $(200\times200)$ for the rest of the domain.

The piece-wise parabolic method \citep{2014JCoPh.270..784M} is used for spatial integration with a $3^{\rm rd}$-order Runge–Kutta
scheme for time evolution and HLL Riemann solver.
For the magnetic field evolution, we apply the DIV-CLEANING
a method that enforces the solenoidal Condition of the magnetic field $\nabla .B =0$

\subsection{Boundary Conditions}

Figure \ref{fig:boundary-condition} shows the numerical setup for our simulations. The left panel shows our axisymmetric computational domain and the corresponding boundary conditions. The right panel, on the other hand, provides a schematic representation of the 3D setup, which includes the boundary conditions and the orbital plane of the binary star. In this section, we consider the left panel which includes the four boundaries in cylindrical coordinates $(r,z)$. At the upper vertical and outer radial directions, we use the standard PLUTO outflow boundary conditions (zero gradients), while on the rotation axis, we apply axisymmetry with $\partial/\partial{\phi} = 0$ (This condition is applied only for our axisymmetric simulation and not for our 3D setup (see section 4.1).

\subsubsection{Equatorial Plane boundary}
\label{disk-boundary}
At the lower part of the vertical direction, we define the ``disk boundary''  that emulates the accretion disk and has an essential role in our model setup.
At the disk boundary, the disk parameters are given as the inflow values \citep{1997ApJ...482..712O, 2011ApJ...742...56V}.

The disk boundary contains the inner gap of the area $ (r < r_{\rm i}) $ and the disk region of $ (r > r_{\rm i})$.
At the ``disk boundary'' the inflow parameters of the sub-keplerian disk are given which should be in equilibrium with the hydrostatic corona initially.
The density component is prescribed by,
\begin{equation}
	\rho_{{\rm disk}(\mathrm{r}, \mathrm{z})}=\left(\frac{1}{1-\chi}\right)\left(\left(\mathrm{r}+\mathrm{r}_{\mathrm{g}}\right)^2+\left(\mathrm{z}+\mathrm{z}_{\mathrm{g}}\right)^2\right)^{-\frac{3}{4}}
\end{equation}
and following the polytrophic law the pressure of the disk material is,
\begin{equation}
 p_{\rm disk}=k \rho_{\rm disk} ^{\gamma}.
 \end{equation}
Here $\chi$ is the Fermi function \citep{2011ApJ...742...56V} and is defined as follows,
\begin{equation}
	\chi=\frac{\chi_0}{1+e^{-10(\mathrm{r}-1)}}.
\end{equation}

Applying the Fermi function imposes a slightly sub-keplerian velocity to fulfill the radial equilibrium at the disk boundary.
The toroidal velocity at the disk boundary is given by,

  \begin{equation}
 v_{\phi}=v_{\rm Kep} \sqrt{\chi} \text \quad { and } \quad
 	v_{\rm Kep}=\frac{1}{\sqrt{r}}.
 \end{equation}
 
Here $v_{\rm Kep}$ is the toroidal velocity at the inner disk boundary $v_{\rm kep}=v_{\phi}(r_{\rm i})$.

{
It should be noticed that we implement time-dependent boundary conditions for the disk boundary, which means that Equation \ref{eta_injection} is used in our setup to determine the poloidal velocity. Consequently, the velocity of the inflowing material at the disk boundary aligns with the poloidal magnetic field ($v_{\rm p} \| B_{\rm p}$) and depends on the evolution of the poloidal magnetic field. This approach differs from previous studies \citep{1997ApJ...482..712O,2011ApJ...742...56V} where fixed-in-time conditions are applied for the outflow inlet.
}

The Stationary axisymmetric MHD flows conserve some quantities
\citep{2007prpl.conf..277P} along the magnetic flux function which is defined by $\psi=r A_{\phi}$ .
For instance, the conservation of mass and the magnetic flux along a field line gives rise to the relation between the poloidal mass flux and the poloidal magnetic field, i.e.,
\begin{equation}
 \eta =\frac{\rho v_{\rm p}}{B_{\rm p}}
 \label{eta_injection}
\end{equation}

where $\eta$ is called the mass load along the field lines and is one of the conserved quantities of the jet.

In addition, the toroidal field in the rotating flows like the jet follows the induction equation which is required to satisfy the following equation,

\begin{equation}
v_{\phi}=\frac{\eta}{\rho}B_{\phi} +r\Omega^{\rm F}.
\label{vphi_constraint}
\end{equation}

Here $\Omega^{\rm F} $ is called the iso-rotation parameter. 
This parameter is interpreted as the angular velocity of the field lines and is another conserved value along the field lines in the jet.
In our set up $\Omega^F$ follows the Keplerian profile and we have
${\Omega^F(r)} \propto r^{-1.5}$.
To prescribe the consistent steady MHD jet, we impose the condition \ref{vphi_constraint} to the disk boundary.

{In the Equatorial boundary, we impose the ideal MHD approximation i.e., $E_{\phi}=(\vec v\times \vec B)_{\phi}=0$ which implies that the poloidal velocity and poloidal magnetic field are parallel, $v_{\rm p} || B_{\rm p}$.
We impose this condition for the vertical magnetic field $B_{\rm z}$ while we use the zero gradient for the radial magnetic field.}

{ We have presented and discussed our (2.5D) axisymmetric simulation in the appendix to examine the 3D simulation of the jet launched in a single star system  which can then be incorporated to non-axisymmetric 3D simulations of the jet in a binary star system.}

\section{ 3D MHD simulation of jet launching }
\label{3Drun_approach}
 In principle, axisymmetry imposes a symmetry constraint that may result in the loss of certain three-dimensional effects in our approach. Consequently, to investigate the formation of a magnetized jet in a binary system, it is necessary to enhance our axisymmetric model setup to incorporate three dimensions and accurately capture the full three-dimensional evolution of the magneto hydrodynamic (MHD) jet launching from a disk boundary.
Therefore, in this section, we present and discuss our 3D reference run, which serves as a basis for our analysis. Subsequently, we implement the companion star into the model and examine the launching of the jet in a binary star system.

\begin{figure*}
    \centering
	\includegraphics[width=19cm]{\figurepath/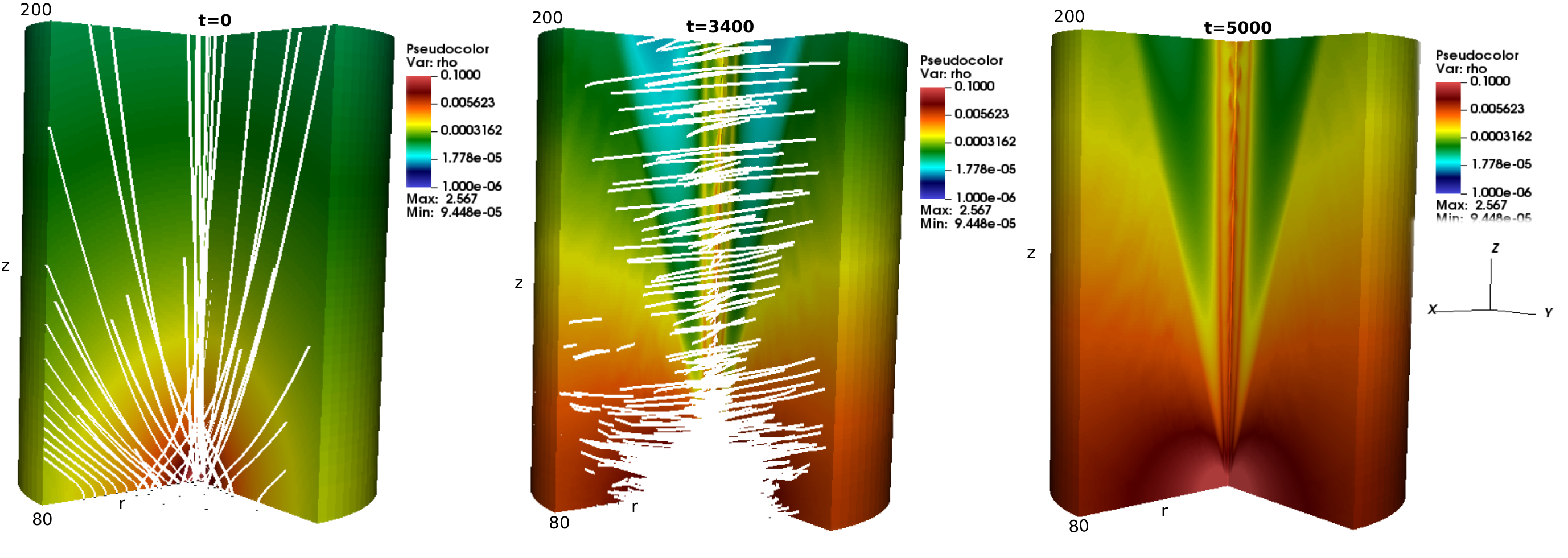}
	\caption{ 3D reference run evolution. Shown are the 3D view of the mass density maps of run "3Dc0" listed in Table\ref{tbl:0} at dynamical times $t=0, 3400, 5000 t_{\rm i}.$
	The overlaid lines represent the magnetic field lines.}
	\label{fig:3D_rho_single}
	 \centering
	\includegraphics[width=18cm]{\figurepath/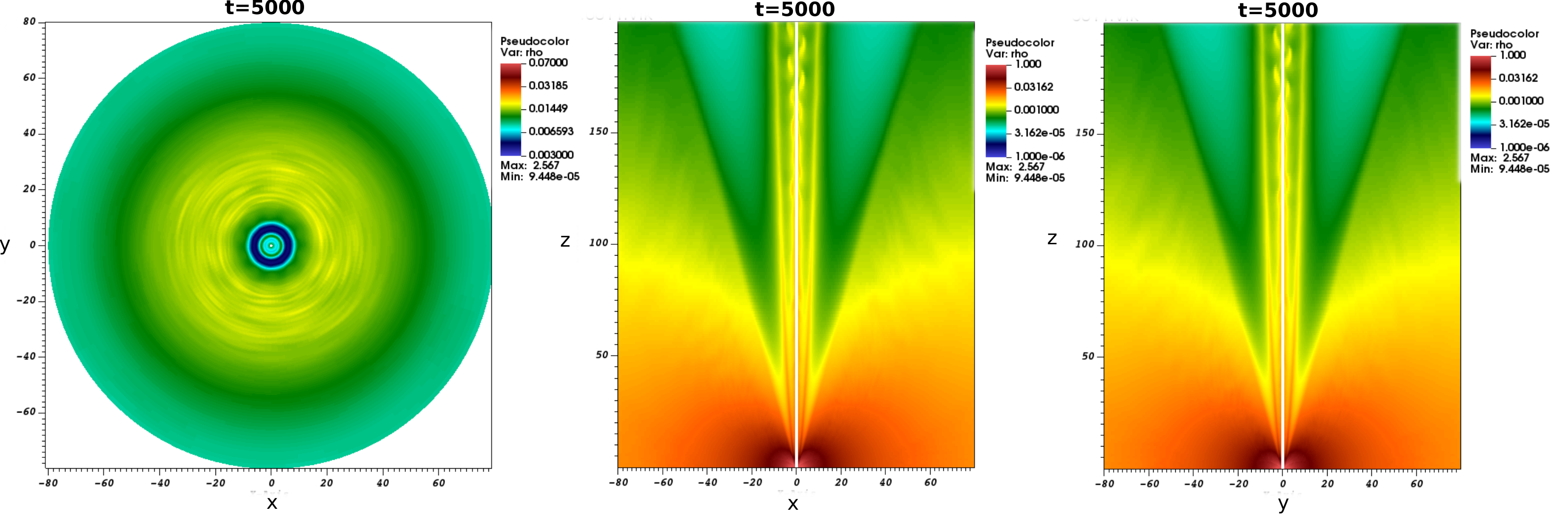}
	\caption{ 3D reference run evolution 2. The displayed images represent 2D slices of the mass density maps at various planes for the "3Dc0" run, taken at dynamical times of $5000 t_{\rm i}$.}
	\label{fig:3D_rho_single_slices}
\includegraphics[width=19cm]{\figurepath/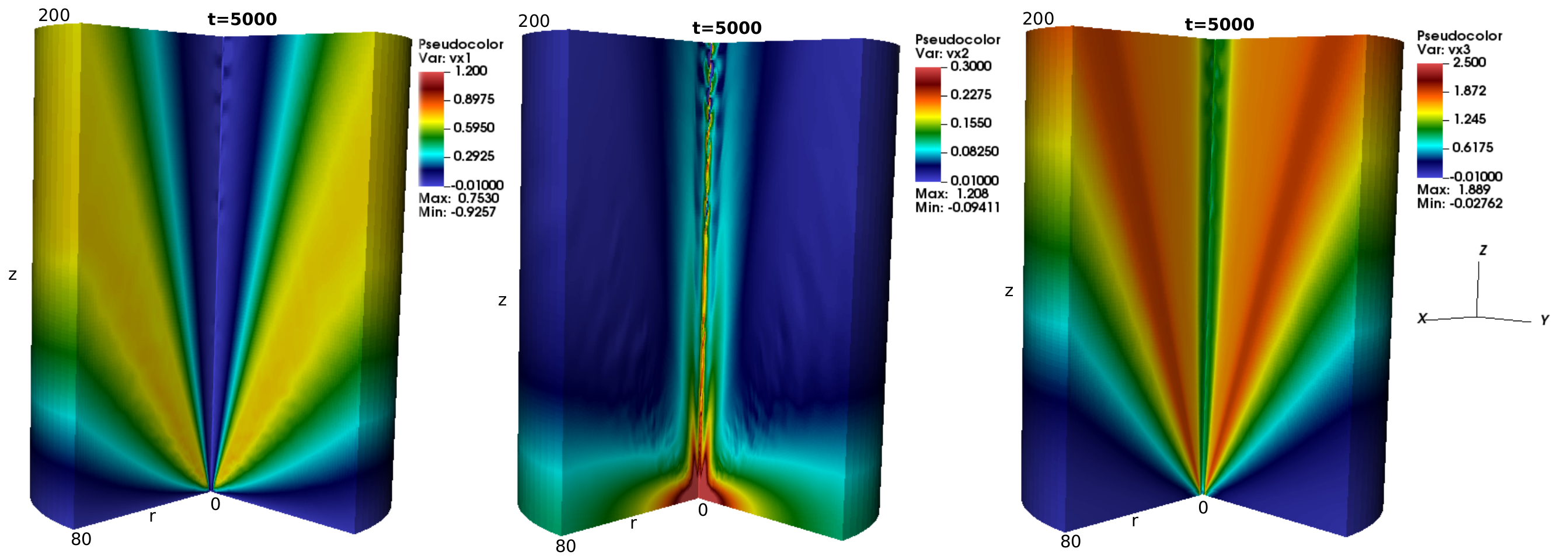}
	\caption{ The variables of the 3D reference run. Shown are the 3D views of the different velocity components of run ``3Dc0'', including the radial velocity component ($v_{\rm r}$), toroidal ($v_{\phi}$), and vertical velocity components ($v_{\rm z}$) of the formed jet, respectively, at a dynamical time of t=5000.}
	\label{fig:3D_velocity_single}
\end{figure*} 

\subsection{3D model setup}
\label{3D Boundary condition}
To enhance our study and extend the model setup to 3D, we utilize polar coordinates $(r, \phi, z)$, as depicted in Figure \ref{fig:boundary-condition} (right panel). The computational domain in our 3D simulation spans from $(0.5,80)$ radially and $(0,200)$ vertically, measured in units of the inner disk radius $r_{\rm i}$. The azimuthal direction covers an angle range of $(0,2\pi)$. Our computational grid consists of $500 \times 100 \times 600$ cells, with a uniform grid in the $\phi$ direction, as well as in the inner radial and vertical directions. Specifically, we incorporate 400 uniform cells for $0 < r < 20$ and 400 uniform cells for $0 < z < 20$, while applying a stretched grid for the remaining part of the grid.

{ It is important to note that, unlike in our axisymmetric  simulation, the $z$-axis no longer serves as a symmetry axis.
Instead, we implement an "inner boundary" (see Figure \ref{fig:boundary-condition}). At the inner boundary which is located at $r=0.5 r_{\rm i}$, we apply a zero gradient for all variables.
{ However, we restrict the velocities to allow for``only'' the inward motion of material into the jet domain. Specifically, we ``only'' allow for negative radial velocity $(v_r <=0)$ and positive vertical velocity (i.e., in a situation where $v_{\rm r}$ is indeed positive, it is set to 0).}

 Applying the zero gradient is a proper choice since the simulation is stable.
We allow for the freely copying  the variables into the ghost cells.
In addition, the inner opening angle is really small and is about 1 degree in our setup.
 If the inner cone is sufficiently small, We would not expect a physical effect, as the flow can always flow besides the axis. Only material that actually hits the inner 1 degree or less in the cone, is lost from the grid. However, it does not affect the flow, since it is not injected again.
 Thus, the physical evolution of the jet is not affected.

For other variables, we utilize a zero gradient condition, meaning that $\partial/\partial z = 0$ and $\partial/\partial r = 0$ at the $z$-axis is applied.
At the outer radial and vertical boundaries, we apply a modified outflow boundary condition that only allows outward radial and vertical velocities. The lower boundary in the $z$ direction represents the disk boundary, as explained in section \ref{disk-boundary}. In the $\phi$ direction, we apply a periodic boundary condition. We utilize the same initial conditions and normalization as our axisymmetric model setup.

\subsection{3D Simulations of Jet Formation in a Single Star System}

In this section, we first discuss our 3D reference run, which presents the formation of a magnetized jet from a disk boundary.
The simulation run time is about 5000 dynamical times, which is long enough for the jet to reach steady states. Figure \ref{fig:3D_rho_single} displays the 3D views of the mass density distribution for run ``3Dc0'' at dynamical times t=0, 3400, 5000 $t_{\rm i}$.
The solid lines represent the magnetic field lines. 
Figure \ref{fig:3D_rho_single_slices} also demonstrates the 2D slices of the mass density maps at various planes at dynamical time of $5000 t_{\rm i}$.

Considering Figure \ref{fig:3D_rho_single} and \ref{fig:3D_rho_single_slices}, it can be observed that a smooth jet is continuously formed from the disk boundary and expands into a larger area.

 \begin{figure*}
\includegraphics[width=17cm]{\figurepath/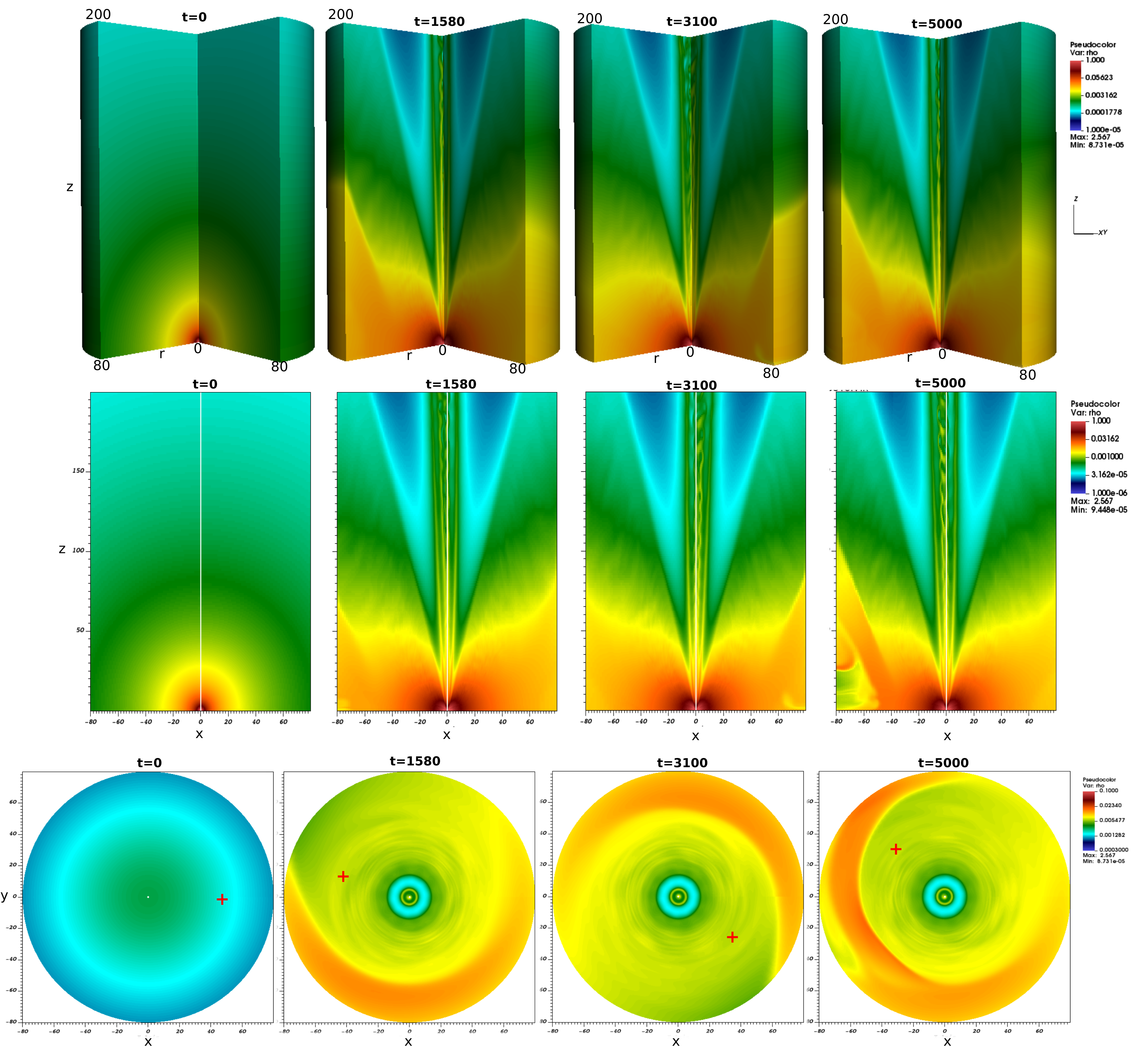}
	\caption{ Jet evolution in a binary system. Shown are 3D rendering and the snapshots of the mass density for run ``3Dc3'' with a mass ratio of the secondary to primary star $q=2$ and the separation of $D=100r_{\rm i}$. The snapshots are shown at different planes and at times 0, 1580, 3100, 5000 $t_{\rm i}$. Here the ``+'' symbol shows the projected position of the inner Lagrange point L1.}
	\label{fig:2Dslices_rho_runbin2_q2}
 \end{figure*}
 
 \begin{figure*}
\includegraphics[width=18cm]{\figurepath/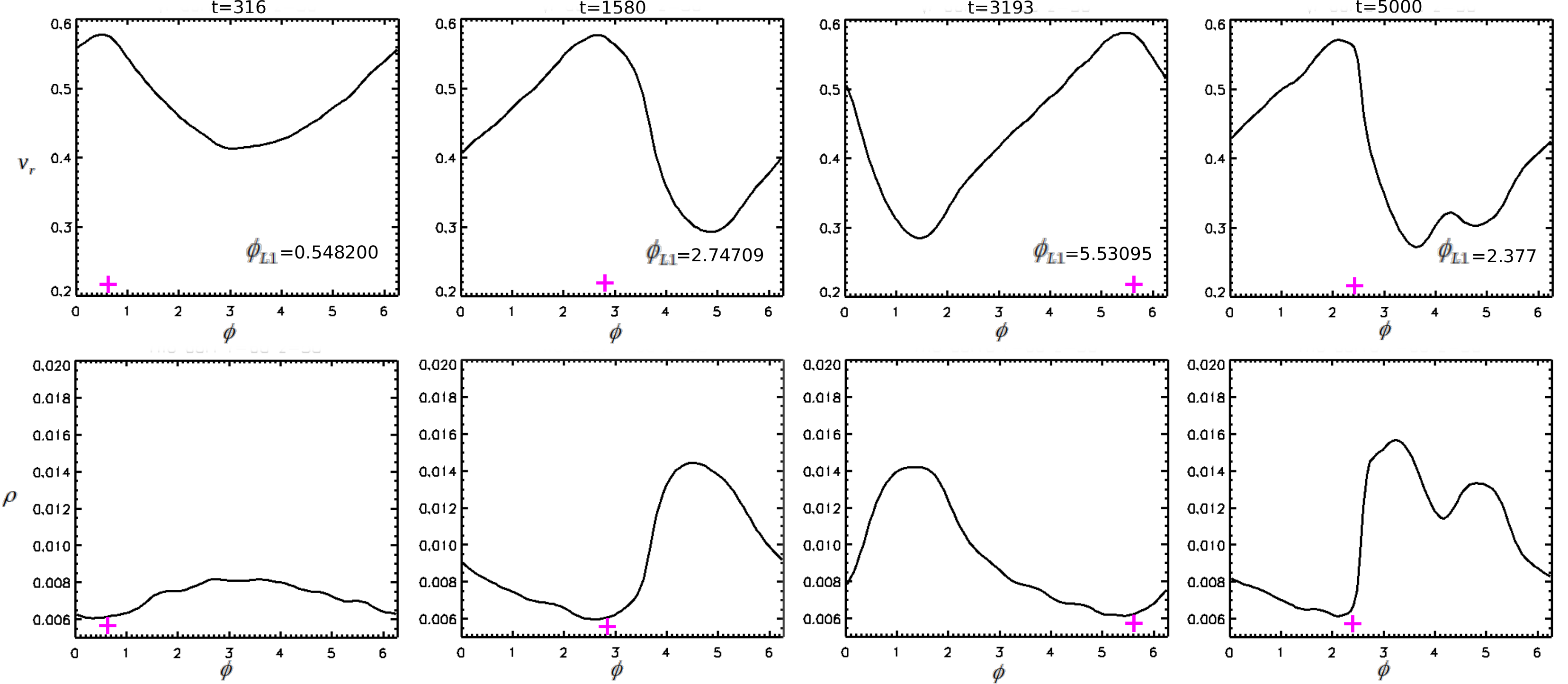}
	\caption{ Angular profile. Shown are the angular profiles of the radial velocity $v_r$ and mass density $\rho$ for the jet in run '3Dc3' with the mass ratio of 2, at radius $r=60$ and $z=50$ and at times t = 316, 1580, 3193, 5000 $t_0$. Here the ``+'' symbol shows the angular position of the inner Lagrange point $(\phi_{\rm L_1})$.}
	\label{fig:runbin2_profile_along_phi}
\end{figure*}

Other physical variables, such as the velocity components, can also be studied to assess the quality of the 3D model setup. Figure \ref{fig:3D_velocity_single} shows the different velocity components of run ``3Dc0'', including the radial velocity component ($v_{\rm r}$), toroidal ($v_{\phi}$), and vertical velocity components ($v_{\rm z}$) of the formed jet, respectively.
We observe that the jet reaches high velocities of about 20 times the local sound speed or about 2 times the inner radius Keplerian velocity, i.e., about $200 km/s$.
The early evolution of the jet is quite fast and the initial shock wave reaches the upper boundary quickly.  
This can be estimated by considering the kinematic timescale of the jet, which is given by the ratio of the distance traveled ($\Delta L$) to the jet velocity ($v_{\rm jet} \simeq 2 v_{\rm i}$). In this case, we obtain $t_{\rm kin} \approx \Delta L / v_{\rm jet} \approx 200 / 2 = 100$.

Considering all the physical variables of the jet in run ``3Dc0'' with the single star, the evolution of the jet material exhibits that the jet keeps the axial symmetry properly, although we do not use the axisymmetric boundary and this shows the quality of the 3D model setup.

  \begin{figure*}
\includegraphics[width=18cm]{\figurepath/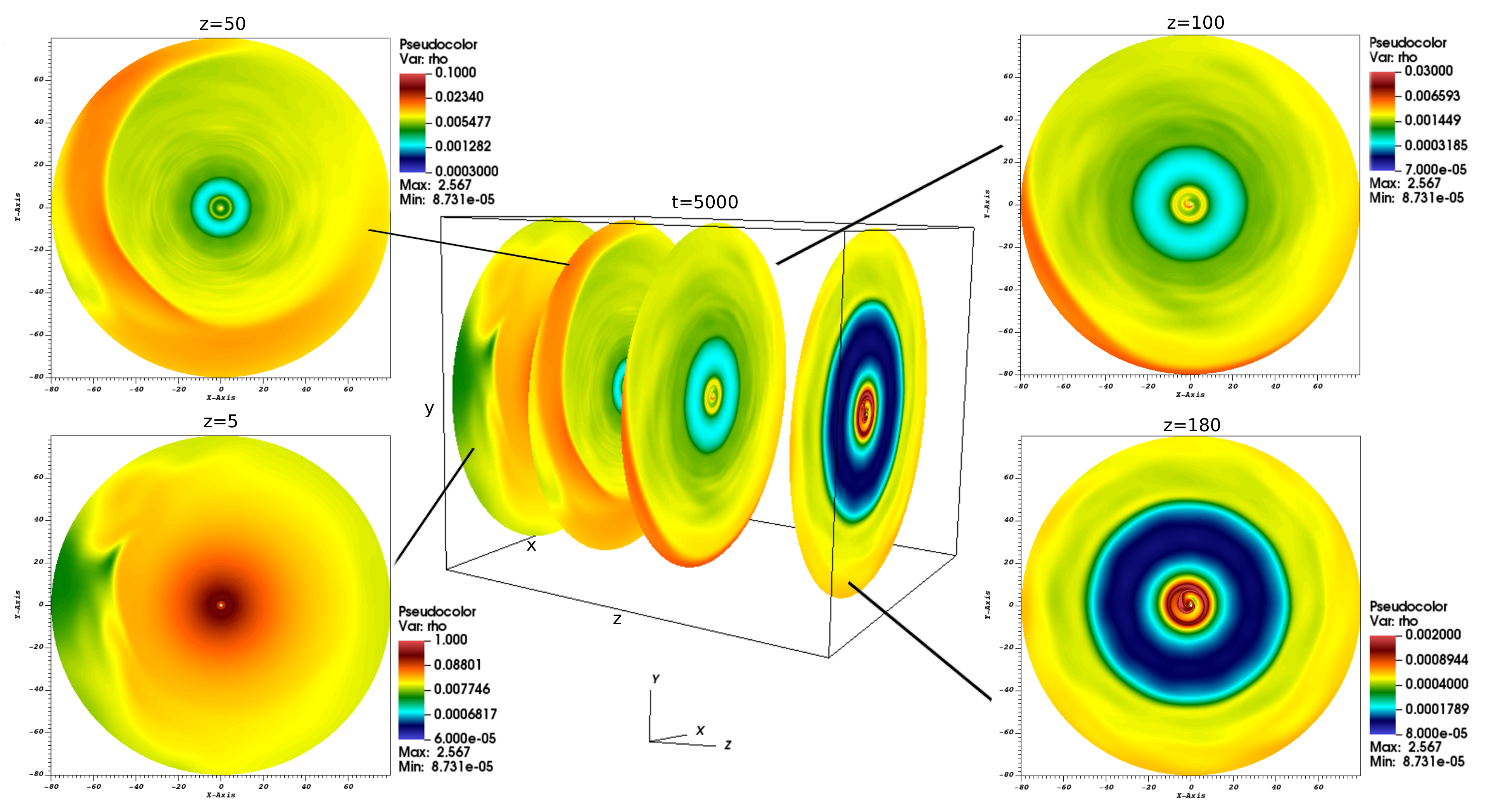}
	\caption{ Jet evolution in binary system. Shown are different snapshots of the mass density for run ``3Dc3'' with the mass ratio of the secondary to primary star $q=2$ and the separation of $D=100r_{\rm i}$. The snapshots are shown at different cross-sections of the jet at z=5, 50, 100, 180 .}
	\label{fig:acrossjet_q2}
 \end{figure*}
\begin{figure*}

\includegraphics[width=17cm]{\figurepath/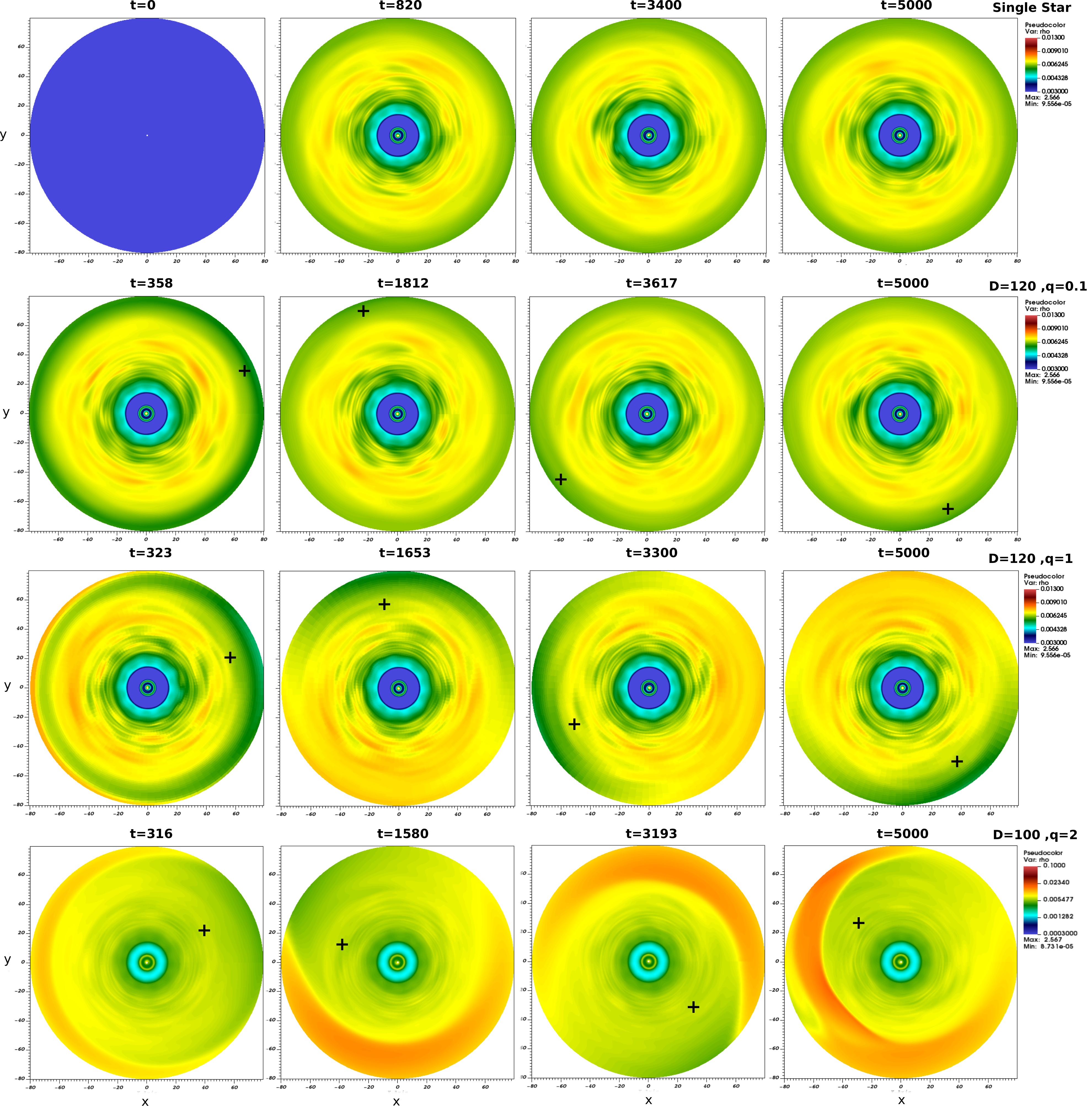}
\caption{ Comparison of jets in binary and single star systems. The snapshots depict the mass density at a cross-section of z=50 for the jets formed in a single star system and a binary system with mass ratios of 0.1, 1, and 2 (from top to bottom) respectively at various dynamical times.
The "+" symbol determines the position of the inner Lagrange point L1 of the binary systems.}
\label{fig:3D_rho_comparision_z50}
\end{figure*}

\subsection{ 3D simulations of jet formation in a binary system}
\label{3D binary}
In this section, we present additional simulation runs that are detailed in Table \ref{tbl:0}. These runs incorporate the Roche potential generated by the companion star. Table \ref{tbl:0} provides the input parameters for our simulations. The key parameters of the binary system include the mass ratio, represented as $q=M_{\rm p}/M_{\rm s}$, which signifies the mass of the secondary star relative to that of the primary star. Additionally, since the origin of the coordinate system remains at the primary star, the binary separation is determined by the position of the secondary star which stays outside the computational domain.

Figure \ref{fig:boundary-condition} (right panel) displays the 3D computational domain and the position of the companion star relative to the primary star. In our model, the primary star is situated at the origin of the coordinate system, while the companion star remains outside the domain at a distance of $r_{\rm s}$. To account for the gravitational effects of the binary system, we utilize an effective potential given by:

\begin{equation}
\Phi_{\rm eff} = - \frac{G M_{\rm p}}{|\vec r|} - \frac{G M_{\rm s}}{|\vec{r}-\vec{D}|} + \frac{G M_{\rm s}}{|\vec{D}|^3}  \left(\vec{r} \cdot \vec{D}\right),
\label{eq:phi_eff}
\end{equation}

where $M_{\rm p}$ and $M_{\rm s}$ denote the masses of the central and companion stars, respectively.
Here $D$ represents the position of the companion star and is defined by:

\begin{equation}
\vec{D}= \hat{x} D \cos{\omega t} + 
         \hat{y} D \sin{\omega t}\cos{\delta} + 
         \hat{z} D \sin{\omega t}\sin{\delta}.
         \label{roche_potential}
\end{equation}
 
The first two terms in the equation \ref{eq:phi_eff} account for the gravitational effects of the primary and companion stars, respectively. The third term is known as the indirect term and arises due to the non-inertial nature of the coordinate system (The origin of our coordinates system stays in the primary star). This term accounts for the acceleration of the origin of the coordinate system and is essential for accurately modeling the motion of the binary system \citep{2018ApJ...861...11S}. 
The parameter $\delta$ shows the inclination angle of the binary orbit concerning the disk boundary. In all the simulations described in this paper, we have set the inclination angle to zero. This means that the binary orbit and the disk boundary are co-planar. It is important to note that the companion star is located outside the computational domain and is not directly included in the simulation.

We have performed the additional runs in three dimensions, referred to as '3Dc1', '3Dc2', and '3Dc3', with mass ratios of the secondary star to the primary  being 0.1, 1, and 2, respectively. To reduce the orbital period of the secondary object and to enhance the tidal effects during the run time, we choose a separation of approximately 100 times the inner disk radius ($r_{\rm i}$) between the secondary and primary objects. It is important to note that in all runs, the orbit of the secondary object remains coplanar with the primary object.
Subsequently, we proceed to compare our other simulation runs with the 3D reference run. By doing so, we can examine the impact of the Roche potential on the physical properties of the jet and gain further insight into the dynamics of the jet present in a binary system.

\section{Jet formation in a binary star system with a high mass ratio}

Here, we consider the evolution of the jet in run "3Dc3" including a binary star with a mass ratio of $q=2$ for the secondary star compared to the primary star.
In Figure \ref{fig:2Dslices_rho_runbin2_q2}, we can observe the evolution of the mass density of the jet material for this run. The figure displays 3D rendering and multiple snapshots of the mass density at different planes, namely the $x-z$ plane and the cross-section of the jet at z=50, at different dynamical times. The "+" symbol determines the position of the inner Lagrange point L1 of the binary systems.

Regarding Figure \ref{fig:2Dslices_rho_runbin2_q2}, we find the direct influence of the secondary star on the mass distribution of the jet. It is evident that the mass density distribution does not maintain axial symmetry, and there is an increase in the amount of jet material. The increase of density is directed towards the inner Lagrange point and due to the orbital motion forms an ``arc-like'' structure.

{ The arc-like structure is visible in the cross-section of the jet at z=50.
This structure exhibits rotation which is synchronized with the orbital motion of the secondary star.
To quantitatively observe the orbital motion of the feature caused by the secondary star, we have presented the angular profiles of the radial velocity $v_r$ and mass density  $\rho$ for the jet in run ``3Dc3'' including a binary star with mass ratio of 2. These profiles were obtained at a radius of $r=60$ and $z=50$, and at various time steps, as shown in Figure \ref{fig:runbin2_profile_along_phi}. 

Analyzing the angular profile of the radial velocity, we observe different peaks occurring at different times, coinciding with the position of the inner Lagrange point  $\phi_{L_1}$ (the ``+``symbol). This indicates that the jet material with higher radial velocity leaves the grid more rapidly resulting in the lower density at that region (''+`` symbol in angular profile of density).
In addition, in the cross-section of the jet and at larger angular distances from $\phi_{L_1}$ (indicated by the "+" symbol), an increase in mass density is observed. The enhanced mass density, combined with the orbital motion of the companion, results in the formation of an arc-like structure.
The angular profile of the radial velocity provides further confirmation that the affected regions within the jet by direct tidal effects, are rotating and synchronized with the orbital motion of the secondary. }

It appears that the arc-like feature is predominantly observed at lower altitudes. As we move further out in the jet, the deviation from axial symmetry becomes less pronounced. This observation is supported by considering the jet at different cross-sections. To provide a clearer understanding and visual representation, we present the mass density of the jet at various cross-sections,i.e., z=5, 50, 100, 180, in Figure \ref{fig:acrossjet_q2}. The figure clearly shows that the enhanced density is formed towards the secondary and is visible at all altitudes.
However, as we move farther away from the companion star and reach greater heights, the impact of this effect diminishes, although it still leaves a discernible imprint in all of the displayed cross-sections.

Moreover, the arc-like structure exhibits a consistent azimuthal angle across various cross-sections of the jet, indicating that there is no noticeable delay in the formation of the arc-like structure along the jet.
\begin{figure*}
\includegraphics[width=17cm]{\figurepath/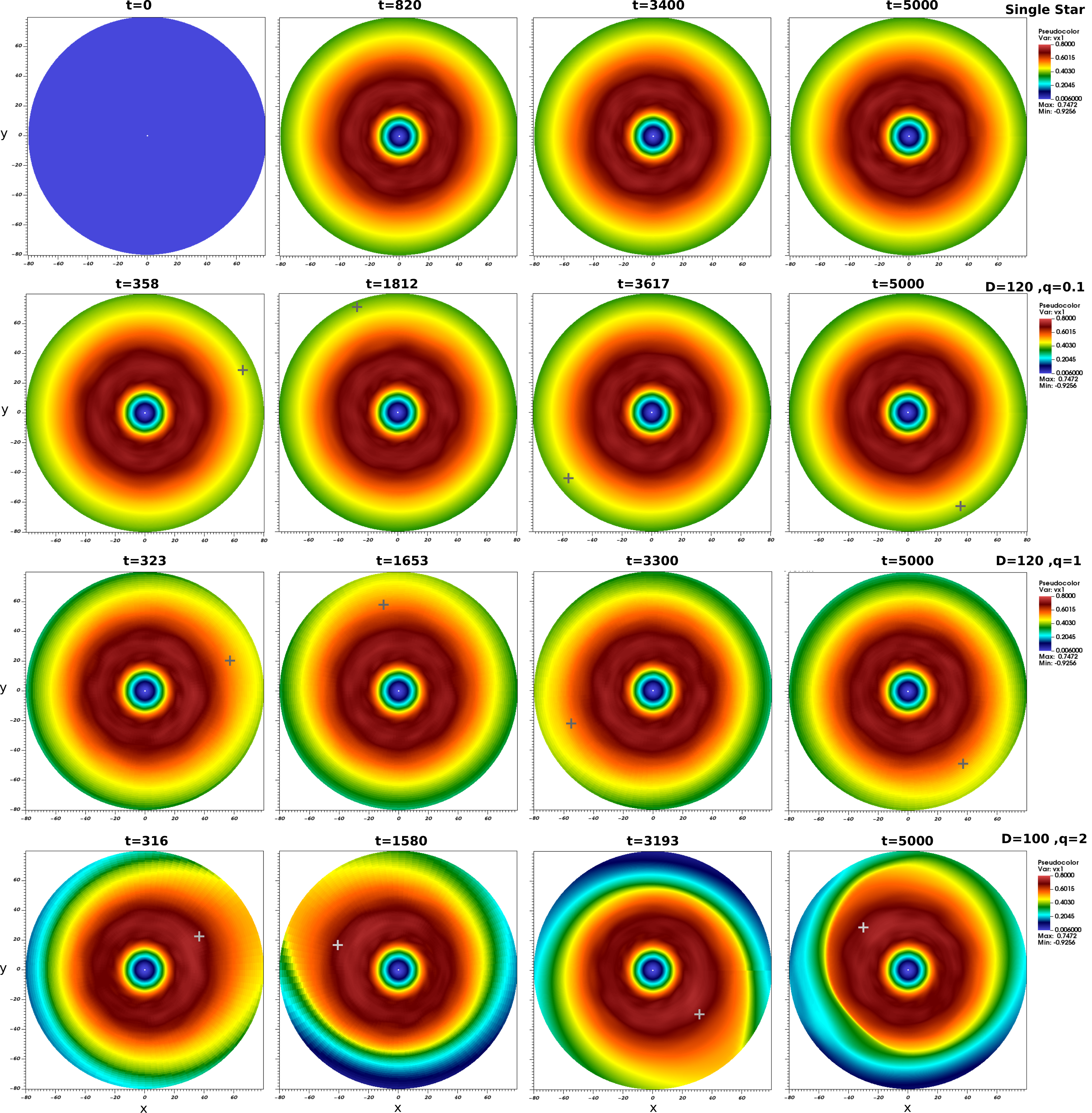}
\caption{ Radial velocity comparison. The provided snapshots display the radial velocity at the cross-section of z=50 for the jets in the runs labeled as '3Dc0','3Dc1, '3Dc2', and '3Dc3', respectively, at various dynamical times. Here the ``+'' symbol shows the position of the inner Lagrange point.}
\label{fig:3D_vr_comparision}
\end{figure*}
\begin{figure}
\centering
\includegraphics[width=0.6\columnwidth]{\figurepath/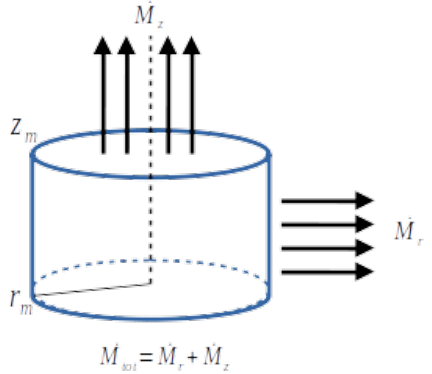}
\caption{ Control volume. Shown is the control volume incorporated to measure the radial and vertical mass flux of the jet. Here $z_{\rm m}$ is chosen to be 100,150,200 and $r_{\rm }$ is 80 (the outer radius of the jet).}
\label{fig:Control_volum}
\end{figure}

\begin{figure*}
\includegraphics[width=18cm]{\figurepath/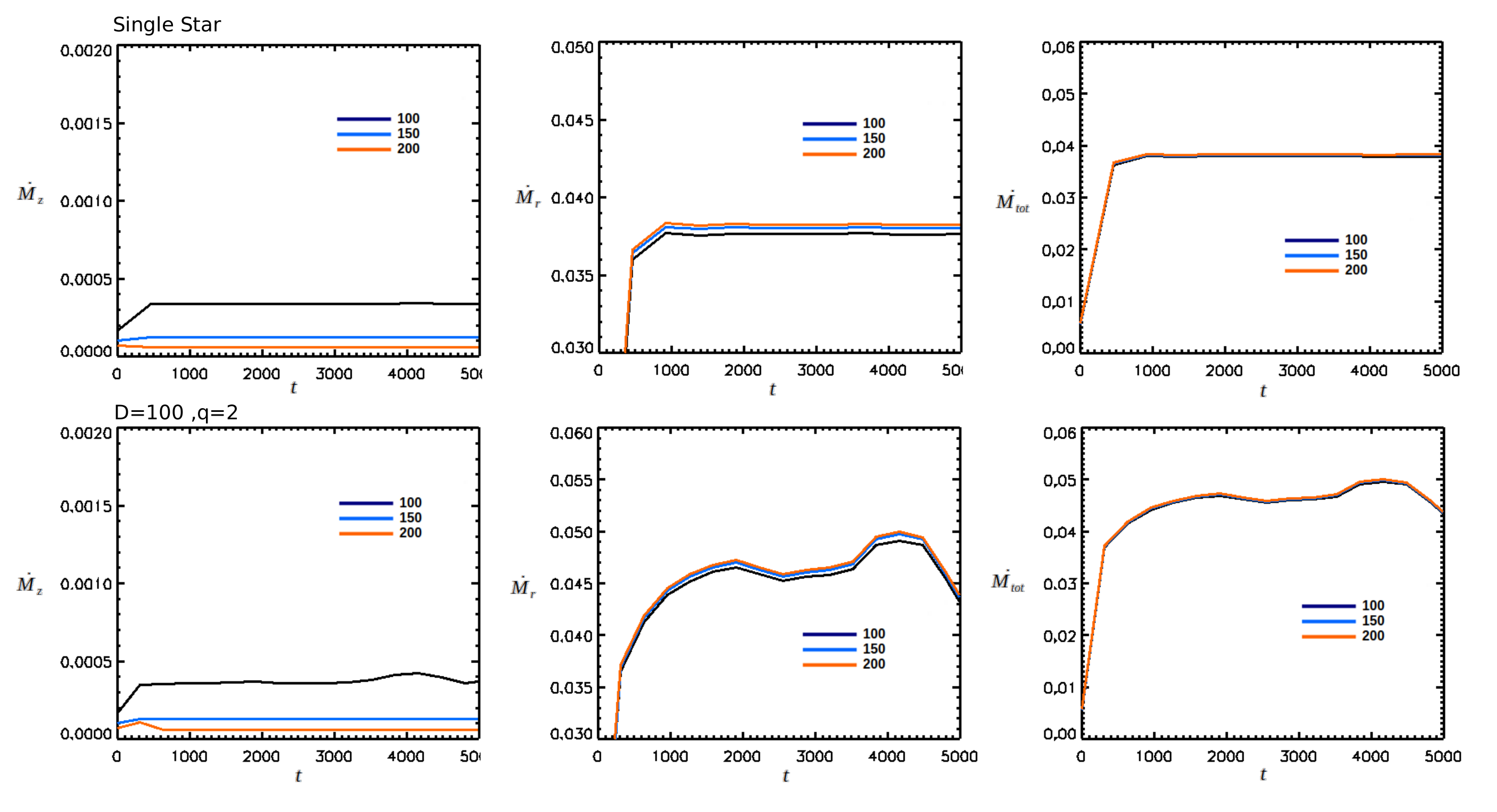}
\caption{ Mass flux conservation. Shown are the time evolution of the vertical $\dot M_{\rm z}$, radial $\dot M_{\rm r}$ and the total mass flux $ \dot M_{\rm tot}$ for control volume at $r_{\rm m}$=80 and $z_{\rm m}$=100,150, 200. The top panels correspond to the jet in single star system and the bottom to the jet in binary star system, with mass ratio of 2.}
\label{fig:mflux_diffz_com}
\end{figure*}
 
\begin{figure*}
\centering
\includegraphics[width=18cm]{\figurepath/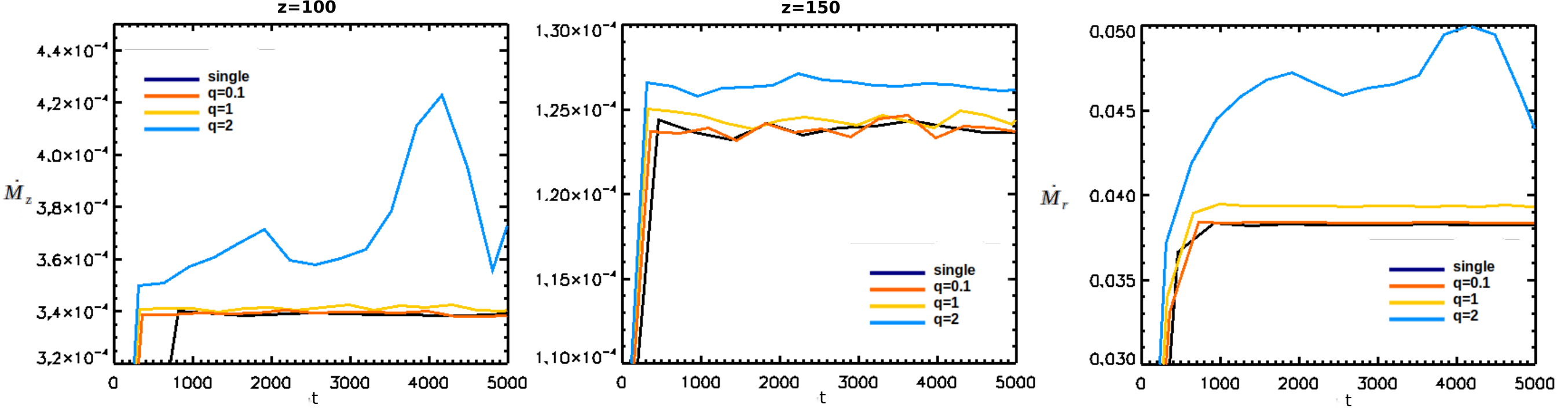}
\caption{Comparison of mass fluxes. Shown are the time evolution of the integrated vertical mass flux of the jet at a height of $z=100r_{\rm i}$ (first) and $z=150r_{\rm i}$ (middle) across the entire cross-section of the jet. The last panel demonstrates the time evolution of the integrated radial mass flux of the jet at a radius of $r=80r_{\rm i}$ across the cylindrical surface, for all 3D runs presented in the paper.}
\label{fig:mfall_all}
\end{figure*}

\subsection{Comparison of all the 3D simulations}

For the sake of the comparison, we have performed different binary runs implementing a secondary with the mass ratios of $q $=  0.1, 1, 2.
Figure \ref{fig:3D_rho_comparision_z50} illustrates the snapshots of the mass density at a cross-section of z=50 for the jets formed in a single star system and a binary system with mass ratios of $q=$ 0.1, 1, and 2 (from top to bottom) respectively at various dynamical times.  The "+" symbol determines the position of the inner Lagrange point L1 of the binary systems.
One can observe that the stronger tidal effects appear due to the larger mass ratio which can produce a denser region in the density map.
It is seen that run ``3Dc1'' with a mass ratio of 0.1 represents a very small deviation from the axial symmetric distribution of the mass density (second row).
In addition, as we increase the mass ratio the deviation becomes larger and a prominent arc-like feature is observed in the jet cross-section of z=50 in run ``3Dc3''.

Additionally, we consider other physical variables, like the radial and toroidal velocity component of the jet, and check if there is any signature of the direct tidal effects is seen.
For instance, Figure \ref{fig:3D_vr_comparision} illustrates the radial velocity of the jet material at the cross-section of z=50 for all the runs listed in Table \ref{tbl:0}. It is observed that the radial velocity disrupts the symmetry and increases towards the inner Lagrange point and the companion star. 
{ In Figure \ref{fig:runbin2_profile_along_phi}, we have demonstrated that the angular position of the peak in radial velocity aligns with the location of the inner Lagrange point, for jet in run `3Dc3''.  Now, Figure \ref{fig:3D_vr_comparision} provides further confirmation that the affected regions within the jet by direct tidal effects, are rotating and synchronized with the orbital motion of the secondary.}

If we specifically examine the radial velocity in the $x-z$ plane for the simulation that incorporates the binary star with a mass ratio of 2, identified as run "3Dc3" in Table \ref{tbl:0}, we can observe a prominent radial deviation of the jet (see Figure \ref{fig:3D_vr_bin_q2} in the appendix). 
This deviation indicates the bending of the jet towards the companion star, which changes over time due to the position of the inner Lagrange point and the motion of the secondary star.
The tidal effects on the jet material are primarily influenced by the extension of the Roche lobe, which, in turn, depends on the binary separation and the mass ratio between the secondary and primary star.
In the vicinity of the Roche lobe, the direct tidal effects are significant, but their effectiveness may diminish on a larger scale. 
Our findings indicate that the companion star exerts significant direct influences, particularly in the region near the Roche lobe of the binary system. These influences are the bending and radial deviation of the jet, as observed in our binary simulations. 

\subsection{Outflow mass flux}

The outflow mass flux is an important observable and it is essential to be measured properly. In principle, what comes in from the disk boundary (inflow boundary condition) must go out from the grids. So if we compare the inflow at the disk boundary with the total outflow mass flux (radial plus vertical mass flux), it must be the same.
 In some previous studies \citep{1997ApJ...482..712O,2002A&A...395.1045F,2010ApJ...709.1100P, 2011ApJ...742...56V}, the disk was taken as a time-independent boundary condition.
However, \citet{1999ApJ...526..631K} applied a more flexible and time-dependent conditions applied for the disk boundary. They have discussed that it is impossible to assign arbitrary fixed values to all the quantities at the disk boundary and some quantities needs to be left free to adjust according to the inflow inlet.
As we mentioned before, in our approach, the disk boundary is defined as a time-dependent and is not fixed in time. In particular, we define the poloidal velocity according to the equation \ref{eta_injection}. Thus in our approach the $v_{\rm p} \| B_{\rm p}$. 
and the poloidal velocity is dependent to the magnetic field evolution as well.
Consequently, the total outflow mass flux evolves during the time and also is dependent on the field strength and distribution. 

In order to check the conservation of the total mass flux (radial plus vertical mass flux) in our simulation, 
we apply different control volumes shown in Figure \ref{fig:Control_volum} where $z_{\rm m}$ is the height of the control volume and $r_{\rm  m}$ is the outer radius of the jet.

We incorporate the control volume and estimate the integrated vertical mass flux of the outflow applying the following integral:
\begin{equation}
	\dot{{M}}_{\rm z} |_{(z=z_{\rm m})}= \int_0^{80 r_{\rm i}} \int_0^{2\pi} \rho {v}_{{z}} r d \phi dr.
\end{equation}
where $\dot{M}_{\rm ejec}$ is the vertical mass flux, $\rho$ is the mass density, $v_z$ is the vertical velocity and $r$ is the cylindrical radius. The integral is taken over the entire cross-sectional area of the jet at $z=z_{\rm m}$ and along the radial extension of $r=0 $ to $r=80r_{\rm i}$. 
Subsequently, we measure the radial mass flux using the following integral:

\begin{equation}
\dot{{M}}_{\rm r} |_{(r=r_{\rm m})}= \int_0^{ z_{\rm m}} \int_0^{2\pi} \rho {v}_{{r}} r d \phi dz.
\end{equation}

Here, $ r_{\rm m}$ indicates the outer radius of the jet and $ z_{\rm m}$ is the height of the control volume.

Figure \ref{fig:mflux_diffz_com} demonstrates the time evolution of the vertical $\dot M_{\rm z}$, radial $\dot M_{\rm r}$ and the total mass flux $\dot M_{\rm tot}$ of the jet in run ''3Dc0`` including the single star (top) and the jet in run ''3Dc3`` including the binary star of mass ratio $q=2$.
The profiles compare the measured mass fluxes for different control volumes of radius $r_{\rm m}=80$ and $z_{\rm m}=100, 150, 200$, respectively. 

It is evident from the figure that the total mass flux measured for different control volume is similar for the jet in single star. In addition, for the jet in binary star system the total mass flux is obtained for different control volumes is similar.
Thus the total mass flux is conserved in each runs.

In addition, it is seen that the total outflow mass flux is slightly larger for the jet in the binary star system compared to the one in single star system. This difference arises due to the evolution of the magnetic field strength and distribution, which deviate from the initial conditions in both runs over time.
In each run the total mass flux is conserved and the time-dependent characteristic of the disk boundary determines the total outflow mass flux.

For example, Figure \ref{fig:BX3compa} illustrates the vertical magnetic field $B_z$ in the $x-z$ plane for the jet in the "3Dc0" run with a single star (top) and the "3Dc3" run with a binary star system having a mass ratio of 2 (bottom) at different dynamical times. The figure demonstrates that the field strength is not symmetric and is larger in certain regions for the jet in run  "3Dc3"  with a mass ratio of 2. 
The stronger magnetic field in the binary system leads to a more efficient magnetic torque and thus a higher outflow mass flux for the jet in binary star. This accounts for the slightly larger total outflow mass flux observed in the binary run compared to the single star run.

We can also compare the radial and vertical outflow mas flux in all runs presented in Table \ref{tbl:0}.

For a better comparison, we estimate the vertical mass flux of the jet at two different heights, $z=100r_{\rm i}$ and $z=150r_{\rm i}$.

Figure \ref{fig:mfall_all} (the first two panels) illustrates the time evolution of the integrated vertical mass flux of the jet at heights of $z=100r_{\rm i}$ and $z=150r_{\rm i}$, encompassing all 3D runs listed in Table \ref{tbl:0}. It is evident from Figure \ref{fig:mfall_all} that the vertical mass flux in the reference run (including the single star) is lower than in the binary runs, confirming that the presence of a secondary star in the binary system affects the mass distribution within the outflow. As we include the binary star and increase the binary mass ratio, the vertical mass flux exhibits an increasing trend, particularly at later dynamical times. This indicates a deviation from axial symmetry in the jet structure, leading to the formation of a dense region inside the jet. Moreover, if we measure the outflow mass flux further out, we observe the same trend, although the overall mass flux is lower than that at z=50 in all runs. This is because the jet material flows out in a poloidal direction (along $\vec v_{\rm p}= \vec v_{\rm r}+ \vec v_{\rm z}$) and it is required to consider the radial mass flux of the outflow, as well.

Figure \ref{fig:mfall_all} (last panel) presents the time evolution of the integrated radial mass flux of the jet at a radius of $r=80r_{\rm i}$ across the cylindrical surface for all 3D runs listed in Table \ref{tbl:0}. We find that the radial mass flux in the reference run is once again lower than in the binary runs, confirming that the mass density and radial velocity change towards the companion star and increase. This radial mass flux is integrated over the cylindrical surface of the jet, indicating the radial deviation of the mass density towards the secondary star. This is another indication on the jet bending caused by the tidal effect on the jet structure.

The obtained trend for the outflow mass fluxes confirms, once again, that while the evolution of the accretion disk plays a significant role in forming the observed features within the jet, the Roche potential also exerts a direct impact on the jet and alters its mass flux.

\subsection{Comparison to our previous study }

In our previous study \citep{2022ApJ...925..161S}, we conducted a comprehensive analysis by simultaneously considering the evolution of both the disk and the jet.
It has been shown that the evolution of the disk in binary systems results in the formation of spiral arms within the disk, which is subsequently propagated into the jet. These spiral arms exhibit a small delay in forming along the jet due to the combined influence of the orbital motion of the binary star and the propagation of the jet.
In addition it has been discussed that on a larger scale, the prominent features observed in the jet structure are attributed to the residual effects from the evolution of the host accretion disk.

In comparison, in the present study, we did not observe the formation of spiral structures within the jet. Furthermore, the region affected by the orbital motion of the binary system displayed a consistent azimuthal angle across different cross-sections of the jet. These findings indicate that the arc-like structure observed in the density or the enhancement of radial velocity is directly induced by the secondary star.
 
Our current results are entirely consistent with our earlier findings in \cite{2022ApJ...925..161S}, which demonstrated that the primary physical effects and features of the outflow at larger scales, such as the formation of spiral arms, originate from the host accretion disk evolution. Additionally, our current study reveals that the direct tidal effects on the dynamics of the jet make a substantial contribution, particularly in the region near the Roche lobe and towards the secondary star.

\section{conclusions}

We have studied the formation of a magnetized jet within a single star and binary star systems. Our specific focus relies on the jet evolution and thus, we treated the disk as a surface and disk evolution is not considered. As a first step, we utilized an axisymmetric run to explore and determine an appropriate model setup that establishes the effective launching of a fast and stable jet from the disk boundary.
Subsequently, we have advanced our setup to ``three dimensions``, enabling us to conduct a comprehensive study of jet formation following the full 3D evolution of a jet launched in single star and binary star systems. 
Thus, we have implemented the companion star (outside the computational domain) and performed various 3D simulations with the mass ratios of $q= 0.1, 1, 2$ to investigate the direct tidal effects on the jet evolution. 

Our study yielded the following results;


\begin{itemize}
\item {We presented the axisymmetric MHD simulations of jet formation from a disk boundary (shown in the appendix). Through our analysis, we find that the global evolution of the jet reaches a steady state. Subsequently, we have enhanced our study and developed our axisymmetric model setup to three dimensions.
Considering our 3D reference run ``3Dc0'', We find that the jet from the disk surface is continuously launched and reaches a high speed of about 20 times the local sound speed which is about the observed values of a typical jet from young stellar objects.}

\item {Furthermore, we find that in our 3D reference  simulation, although  we do not employ an axisymmetric boundary for the z-axis, the formed jet exhibits a reasonable degree of axial symmetry, particularly after the early evolution ($t\simeq$ 300). This observation indicates the quality  of the 3D model setup which can be incorporated for studying the full 3D evolution of the jets launched in a binary system.}

\item{ By implementing the secondary star in the model setup, we find distinct variations in the physical parameters of the jet. For instance, we see the significant difference in the mass density distribution or the radial velocity $v_{\rm r}$ components of the jet in binary star compared to the single star system, Specifically, inside the jet and toward the secondary star an increasing in the mass density is formed and represent an ``arc-like'' structure.
This feature is rotating and is synchronized with the orbital motion of the secondary.
Our findings suggest that the companion star has a substantial and direct influence, especially in the region near the Roche lobe of the binary system. These influences manifest as the formation of a denser feature within the jet, as well as the bending and radial deviation of the jet, as observed in our binary simulations.
}

\item{ Based on our analysis of the total mass flux for the jets in the single star and binary star systems, we can conclude that the total mass flux is conserved within each run when considering different control volumes. However, the time-dependent nature of the disk boundary plays a crucial role in determining the total outflow mass flux.

Furthermore, we have observed that the total outflow mass flux is slightly greater in the jet of the binary star system compared to the single star system. This difference can be attributed to the evolution of the magnetic field strength and distribution, which deviate from their initial conditions as the simulations progress over time. }

\item{
Moreover, based on evaluating the vertical mass flux in all runs, we find that the integrated value in the reference run (single star) is lower than in the binary runs, and this trend increases as the mass ratio increases.
It indicates that the presence of a secondary star, influences the mass distribution and the velocity field, especially at the outer part of the outflow, leading to a larger deviation from the vertical motion of the materials and a redistribution of mass within the jet. 
}

\item{ Our results are entirely consistent with our earlier findings in \cite{2022ApJ...925..161S}, which demonstrated that the primary physical effects and features of the outflow at larger scales, such as the formation of spiral arms, originate from the host accretion disk evolution. Additionally, our current study reveals that the direct tidal effects on the dynamics of the jet make a substantial contribution, particularly in the region near the Roche lobe and towards the secondary star.}
\end{itemize}

We thank Andrea Mignone and the PLUTO team for the possibility of using their code. We would also like to acknowledge the helpful and insightful comments from an anonymous referee, which have greatly contributed to productive discussions and an improved presentation of our findings.
We would also thank Christian Fendt for the helpful comments.
Our simulations were performed on the TURIN cluster of Institute for Research in Fundamental Sciences (IPM) and the COBRA cluster of the Max Planck Society.

\appendix

\section{General evolution of axisymmetric jet launching}
\label{2.D_run}

 Here, we present and discuss the results of our axisymmetric reference run ``2Dc0'' which investigates axisymmetric jet launching.
Figure \ref{fig:rho_reference} shows the time evolution of the mass density, with solid lines indicating the poloidal magnetic field lines. The continuous and powerful jet is seen to form from the lower part of the grid (at the disk boundary) and develop outward into a larger area. As the jet spreads out, it forms a bow shock that affects the corona and sweeps up materials. The resulting jet exhibits typical collimation, which is evident and caused by the toroidal component of the magnetic field. Our model setup and the resulting jet confirm the validity of the "disk-wind" theory originally proposed by \cite{1982MNRAS.199..883B} and later supported by \cite{1983ApJ...274..677P}.

\begin{figure*}
\centering
\includegraphics[width=15cm]{\figurepath/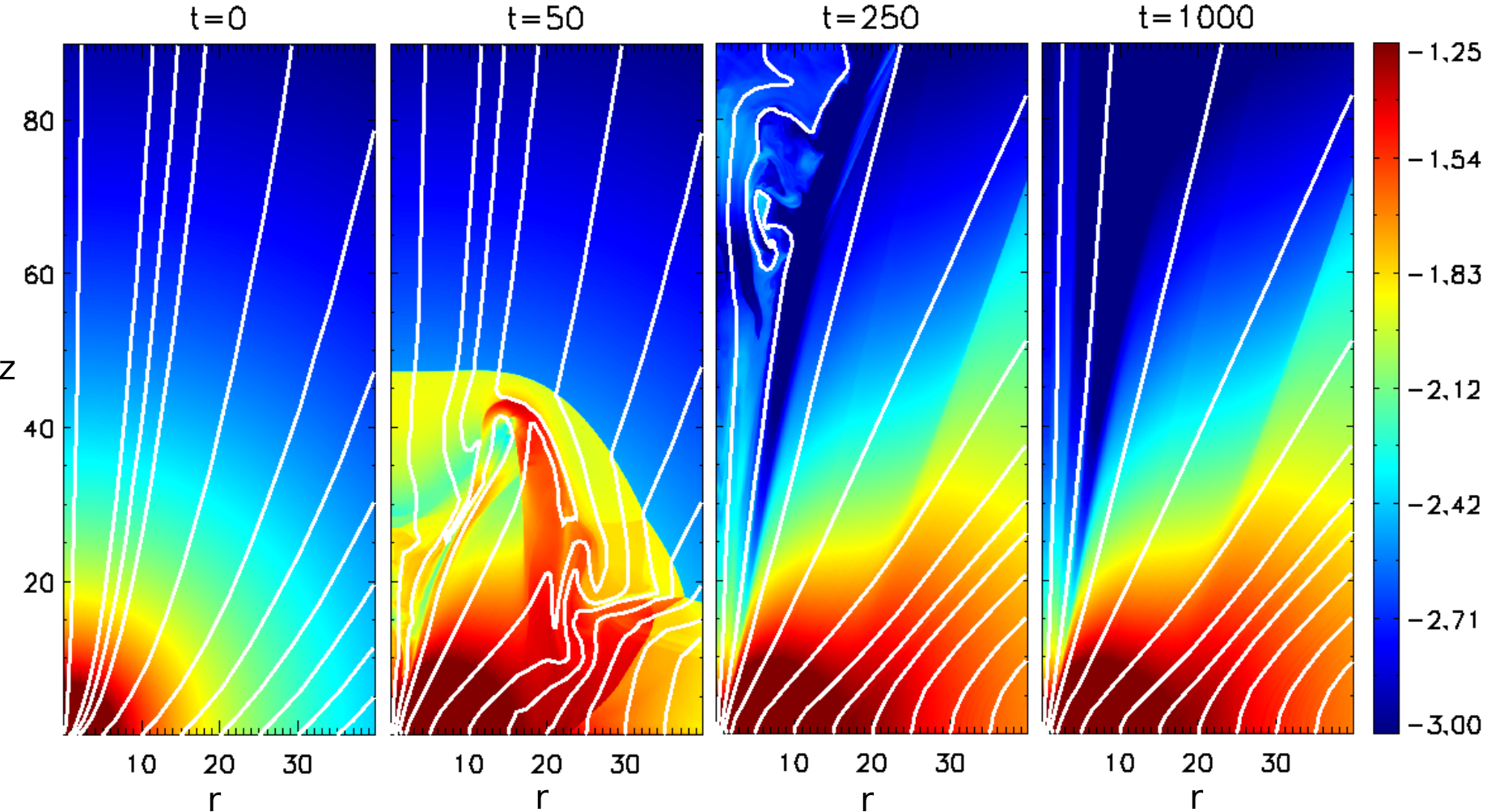}
\caption{Reference run. Shown are the maps of the mass density in the Logarithmic scale for the dynamical times t=0, 50, 100,1000. The solid lines indicate the poloidal magnetic field lines.}
\label{fig:rho_reference}
\end{figure*}
\begin{figure*}
\centering
\includegraphics[width=15cm]{\figurepath/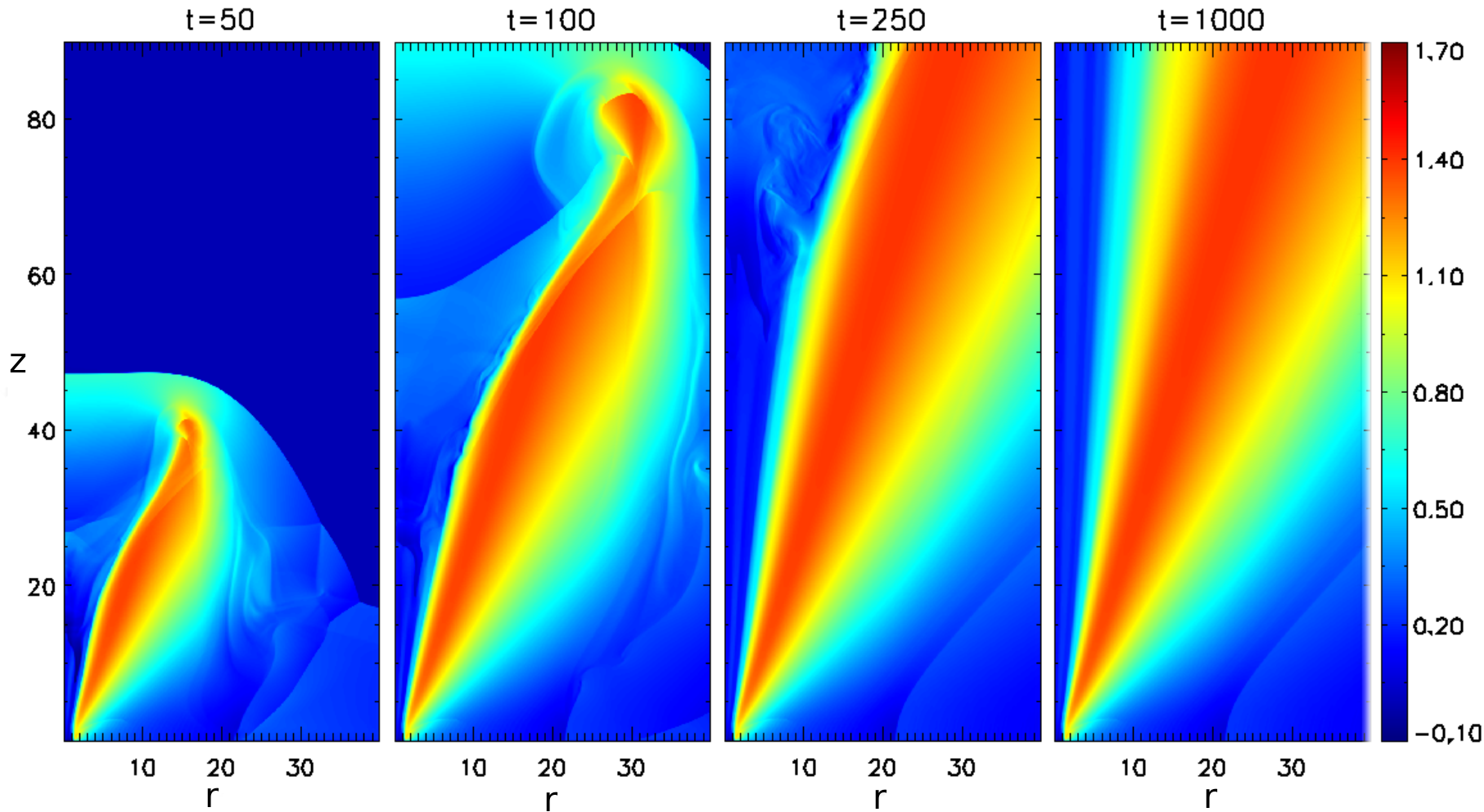}
\caption{Reference run. Shown are the maps of the vertical velocity of the reference run (C1) for the dynamical times, t=50, 100, 250, and 1000.}	
\label{fig: V_Reference}
\end{figure*}

\begin{figure}
\centering
 \includegraphics[width=8cm]{\figurepath/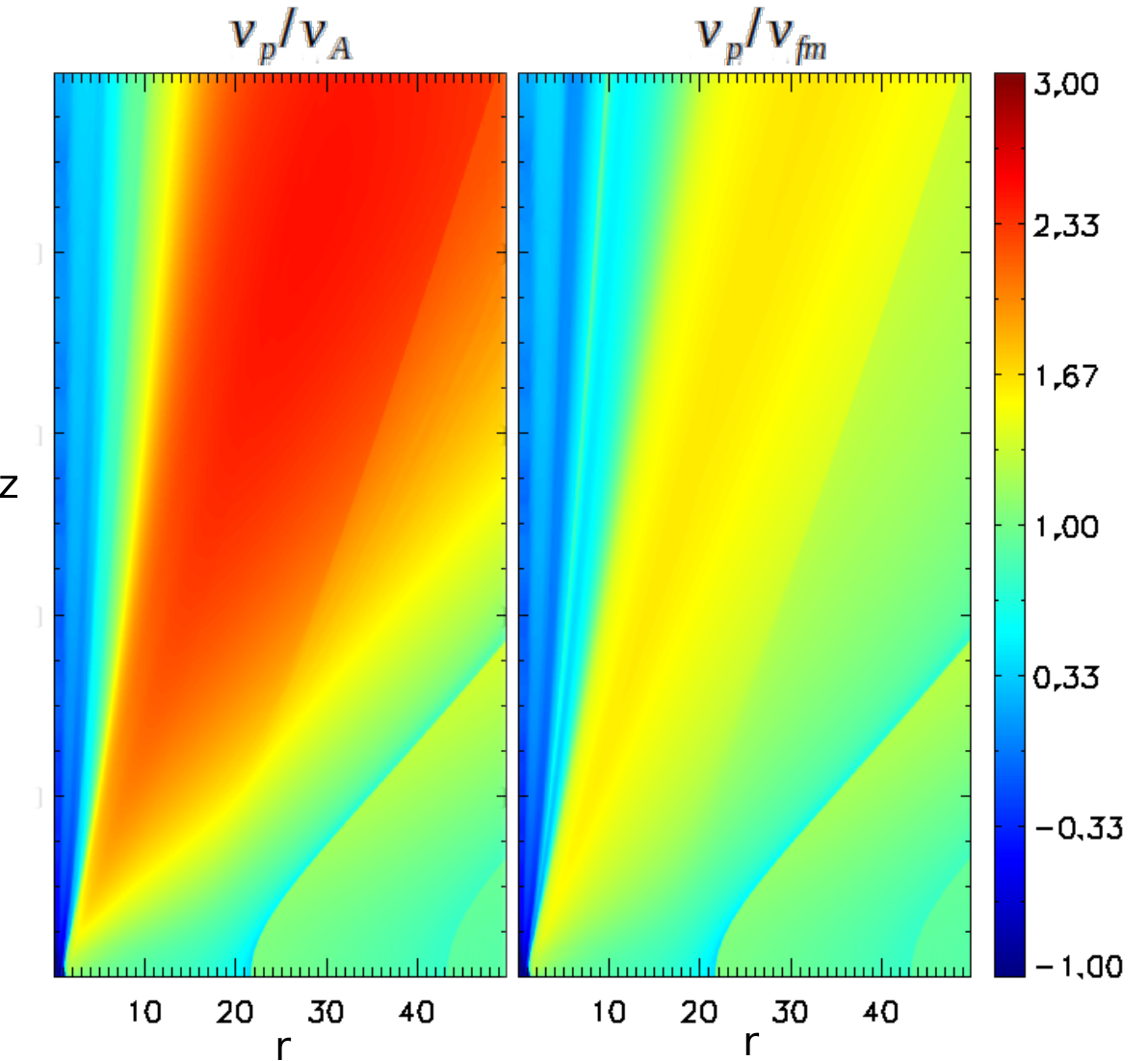}
\caption{Reference run. Shown are the snapshots of the ratio of the poloidal velocity to the Alfven and fast magneto-sonic speed for the reference run in logarithmic scale and at dynamical time 1000, respectively.}
\label{fig:fast_slow_reference}
\end{figure}

Figure \ref{fig: V_Reference} displays the vertical velocity map of the axisymmetric reference run. Considering the figure, we observe that the jet materials initiate with low velocities and achieve high speeds as time passes. The bow shock is visible in the velocity map and eventually leaves the grid.
In addition, Figure \ref{fig:fast_slow_reference} shows the snapshots of the ratio of the poloidal velocity $v_{\rm p}$ to the Alfven $ v_{\rm A}$ and fast magneto-sonic $v_{\rm fm} $ speed for the reference run in logarithmic scale and at dynamical time 1000, respectively.
Referring to Figure \ref{fig:fast_slow_reference}, it is apparent that the outflow materials exhibit highly super-Alfvenic behavior and reach super-fast magneto-sonic speeds.

The global evolution of the  axisymmetric reference run "2Dc0" demonstrates that the jet achieves a steady state, indicating that it can be employed as a basis for extending the model to 3D, in the next step of our study.

\section{The physical variables of a jet in a binary system with a mass ratio of 2}
Here, we present some other variables of the jet in run ``3Dc3'' with the mass ratio of 2.

\begin{figure*}
\includegraphics[width=18cm]{\figurepath/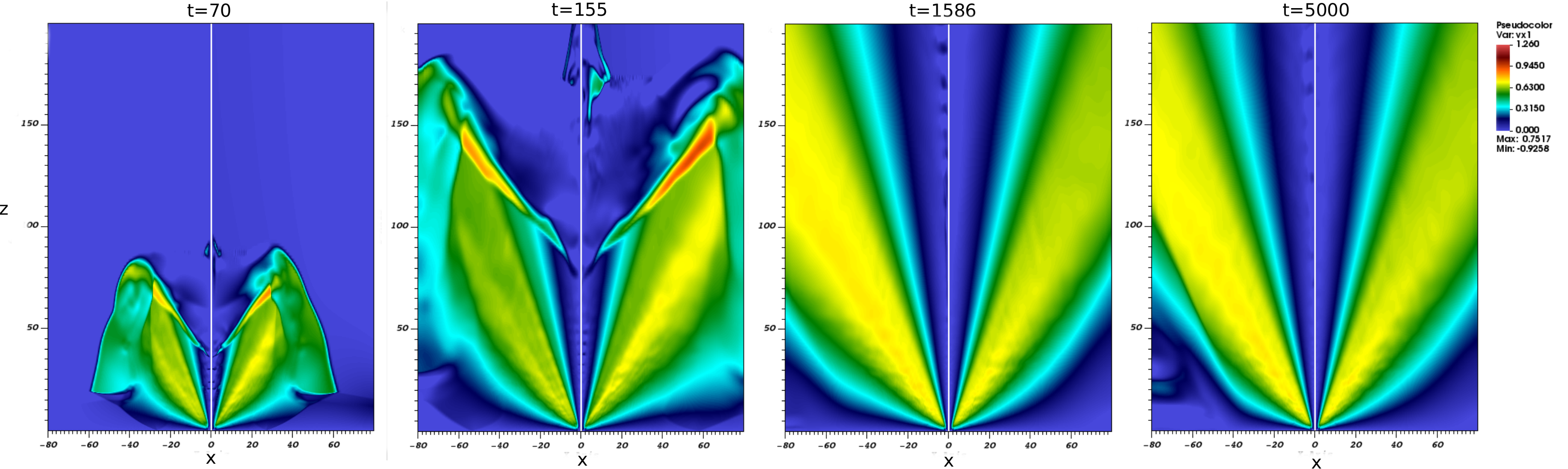}
\caption{ Radial velocity evolution. The provided snapshots display the radial velocity at plane $x-z$ for jet in run ``3Dc3'' with the mass ratio of 2, at various dynamical times.}
\label{fig:3D_vr_bin_q2}
\end{figure*}

\begin{figure*}
\includegraphics[width=18cm]{\figurepath/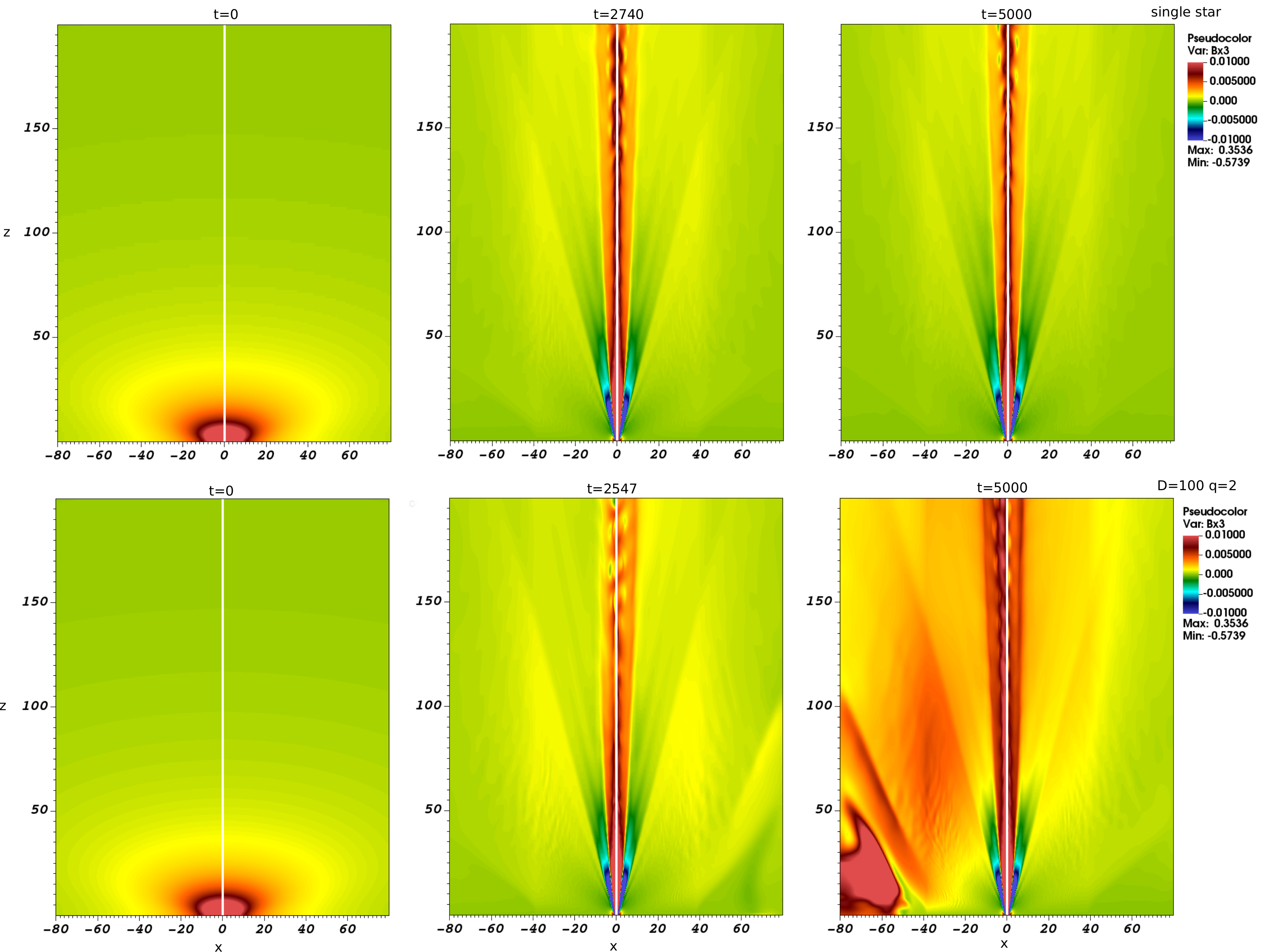}
\caption{ Vertical magnetic field. The presented snapshots depict the vertical magnetic field $B_z$ in the $x-z$ plane for the jet in the "3Dc0" run with a single star (top) and the "3Dc3" run with a binary star system having a mass ratio of 2 (bottom) at different dynamical times.}
\label{fig:BX3compa}
\end{figure*}

\bibliographystyle{apj}
\bibliography{bibpaper2021}

\end{document}